\documentclass[a4paper,USenglish,cleveref, autoref, thm-restate]{lipics-v2021}

\pdfoutput=1 
\hideLIPIcs  

\usepackage[utf8]{inputenc}
\usepackage{graphicx} 
\usepackage{color}
\usepackage{hyperref}
\usepackage{amsmath}    
\usepackage{amsfonts}
\usepackage[noend]{algpseudocode}
\usepackage{algorithm}
\usepackage{dirtytalk}
\usepackage{amsthm}
\usepackage{xspace}

\usepackage{pict2e}
\usepackage{amssymb}

\usepackage{enumerate}

\usepackage{tikz}
\usetikzlibrary{automata, arrows.meta, positioning}

\newboolean{comments}
\setboolean{comments}{false}

\newcommand{\maxence}[1]{{
    \ifthenelse{
        \boolean{comments}
    }{
        \color{red}{{M: #1}}
    }
    {}
}}

\newcommand{\sara}[1]{{
    \ifthenelse{
        \boolean{comments}
    }{
        \color{blue}{{S: #1}}
    }
    {}
}}

\newcommand{\petr}[1]{{
    \ifthenelse{
        \boolean{comments}
    }{
        \color{brown}{{P: #1}}
    }
    {}
}}

\newcommand{\set}[1]{\textit{set(}#1)}

\newcommand{\operations}[0]{{O}}
\newcommand{\dags}[0]{\mathcal{D}}

\newcommand{\Nat}{\mathbb N}
\newcommand{\rdt}{(Q,q_0,\operations,R,\sigma)}
\newcommand{\command}{(o,i,s)}
\newcommand{\commandSet}{(\operations\times\Pi\times\Nat)}
\newcommand{\fbfs}[0]{$f_{\textit{BFS}}$\xspace}
\newcommand{\ffair}[0]{$f_{\textit{fair}}$\xspace}

\newcommand{\seq}{\textit{seq}}
\newcommand{\past}{\textit{past}}
\newcommand{\length}{\textit{length}}
\newcommand{\concurrent}{\textit{concurrent}}

\newcommand{\dist}{\textit{dist}}
\newcommand{\parents}{\textit{parents}}

\newcommand{\hide}[1]{}

\title{Wait-free Replicated Data Types and Fair Reconciliation}

\author{Petr Kuznetsov}{Télécom Paris, Institut Polytechnique de Paris, France}{}{https://orcid.org/0000-0003-1148-1228}{}
\author{Maxence Perion}{Université Paris-Saclay, CEA, List, France}{maxence.perion@cea.fr}{https://orcid.org/0009-0005-4048-6708}{}
\author{Sara Tucci-Piergiovanni}{Université Paris-Saclay, CEA, List, France}{}{https://orcid.org/0000-0001-9738-9021}{}

\authorrunning{P. Kuznetsov, M. Perion, and S. Tucci-Piergiovanni} 

\Copyright{Petr Kuznetsov, Maxence Perion, and Sara Tucci-Piergiovanni} 

\ccsdesc[300]{Theory of computation~Distributed algorithms}
\ccsdesc[300]{Computer systems organization~Dependable and fault-tolerant systems and networks}

\keywords{Replication, Revocation, Eventual consistency, CAP, CRDT, DAGs, Fairness}

\category{} 

\supplementdetails[linktext={github link},subcategory={Source Code}, swhid={swh:1:dir:7f5598679623666993670564cc449edd0b4f97c0;origin=https://github.com/maxper4/fair-wait-free-replicated-data-types-experiments;visit=swh:1:snp:eb182b0c08205dab39b3b05d721e47cd2addf120;anchor=swh:1:rev:e1257a4b09ac88a9bb350844995e1afde3386d0e}]{Software}{https://github.com/maxper4/fair-wait-free-replicated-data-types-experiments/tree/docker-experiments} 

\nolinenumbers 

\EventEditors{Ioannis Chatzigiannakis, Andrea Vitaletti, Keren Censor-Hillel, and William K. Moses Jr.}
\EventNoEds{4}
\EventLongTitle{40th International Symposium on Distributed Computing (DISC 2026)}
\EventShortTitle{DISC 2026}
\EventAcronym{DISC}
\EventYear{2026}
\EventDate{November 9--13, 2026}
\EventLocation{Rome, Italy}
\EventLogo{}
\SeriesVolume{397}
\ArticleNo{32}

\begin{document}

\maketitle

\begin{abstract}
Replication ensures data availability in fault-prone distributed systems. 
The celebrated CAP theorem stipulates that replicas cannot guarantee both strong consistency and availability under network partitions. 
A popular alternative, adopted by CRDTs, is to relax consistency to be \emph{eventual}.
It enables progress to be \emph{wait-free}, as replicas can serve requests immediately, based on their local states. 

Yet, wait-free replication faces a key challenge: due to asynchrony and concurrency, operations may be constantly \emph{reordered}, leading to results inconsistent with their original contexts and preventing them from stabilizing over time. 
Moreover, a particular client may experience \emph{starvation} when, from some point on, \emph{each} of its operations is reordered at least once.
This may lead to pathological scenarios, when the locally observed effect of each of the client's operation is later rejected by the system. 

We make two contributions. 
First, we formalize the problem addressed by wait-free replicated data types (e.g., CRDTs) as \emph{eventual state-machine replication}. 
We then augment it with \emph{stability} and \emph{fairness} ensuring, respectively, that (1)~all replicas share a growing \emph{stable} prefix of operations, and (2)~no client starves.
Second, we present a \emph{generic DAG-based framework} to achieve eventual state-machine replication for any replicated data type, where replicas exchange their local views and merge them using a \emph{reconciliation function}.
We then propose reconciliation functions ensuring stability and fairness. 
\end{abstract}

\section{Introduction}
\label{sec:intro}
Services replicating data or computations over many servers tolerate some of them being temporarily unavaliable or even faulty.
 They maintain multiple copies of the shared data and use \emph{synchronization} protocols to ensure that the evolving replicas are \emph{up-to-date}.
The celebrated CAP theorem~\cite{cap,cap-lynch} stipulates that, if the system is prone to network partitions, \emph{strong consistency} (intuitively, the set of replicas creates the illusion of a unique correct server) and \emph{availability} (intuitively, replicas can serve timely responses for commands) cannot be implemented in the same system.

Availability is often indispensable in practice~\cite{vogels-ec}: clients would feel abandoned if the system responds slowly, affecting negatively their perception of the service.
In the common case when partitions are unavoidable, it is natural to resort to weaker consistency criteria such as \emph{Strong Eventual Consistency (SEC)},  the consistency criterion satisfied by \emph{Conflict-free Replicated Data Types (CRDTs)}~\cite{crdt11}.
Intuitively, SEC allows the states of two replicas to diverge arbitrarily for a period of time, but should \emph{converge} as soon as they received the \emph{same set of commands}.
From this relaxation of consistency, replicas gain \emph{wait-freedom}\footnote{We intentionally use the same word, originally defined in shared memory as \say{processes progress regardless of the speed of other processes}. Lifting these definition to message-passing implies that replicas cannot wait for any message. The definition of wait-freedom we consider is formally defined in Section~\ref{sec:problem}.}~\cite{saito-shapiro-or,Her91}: the local copy of the data may serve to produce a response immediately, without the need to consult with other replicas,
a principle also known as \emph{optimistic replication}~\cite{saito-shapiro-or}.
This approach found numerous applications~\cite{crdt11, soundcloudcrdts, tomtomcrdts, apollocrdts}, commonly used in geo-replicated databases~\cite{redis, riak, crdtmultisyncdb}.

While CRDTs and optimistic replication~\cite{saito-shapiro-or} have been implemented in numerous systems~\cite{soundcloudcrdts,tomtomcrdts,apollocrdts}, we argue that the problem they aim to solve remains poorly defined. 
This lack of formalization has led to a variety of ``CRDT-like" systems providing different guarantees. 
For instance, some implementations are not wait-free, as they rely on some form of coordination (such as quorum acknowledgments) before replying to clients~\cite{riak,BFTgrowOnlySets,antidotedb,processCommutativeObjects}.
Other systems depart from the notion of being strictly conflict-free (which, formally, only holds for data types whose operations commute in all states), by replicating data types with non-commutative updates, handled through an arbitration strategy~\cite{soundcloudcrdts, antidotedb,verifrdt}.
For instance, a popular strategy is \say{remove-wins} for a set data type, where removing an element is preferred to its concurrent addition, effectively ``undoing'' the addition. 

But what happens when a system is \emph{both} wait-free, so fully partition-tolerant, and not strictly conflict-free?

First we observe that if the response to a local command must be provided without consulting other replicas, some of concurrent conflicting commands will have to later be \emph{reordered}, which might change their effect on the shared state and their responses. %
Once the replicas with conflicting commands detect the conflict, they will have to apply some arbitration strategy to decide on the order in which they should be executed.
This way a command originally executed (in the wait-free manner) in a given \emph{context} (a replica state) can later be rexecuted \emph{after} conflicting concurrent commands, thereby changing the context.  

A client therefore receives weak guarantees: over time, the effect of her commands may differ from the effect originally intended in their original contexts.
Reordering may be a result of genuine concurrency or more notably be \emph{deliberately triggered} by a malicious client trying to gain an advantage~\cite{DagWithoutFinality,sandwich}. 
Harmless in purely conflict-free data types, reordering may be problematic in data types with conflicts.
Think of a data type where some operations are enabled only in certain states. 
Consider, e.g., an eventually consistent Network File System~\cite{CRDTforconcurrentfilesystems}, and assume that a client, Alice, creates a folder and then adds a file to that folder (an operation which is only enabled when the folder is present).  
Now assume that another client, Bob, periodically optimizes the shared space by deleting folders that remain empty long enough. 
If Alice is slow or poorly connected, the folder deletion commands of Bob can always be ordered \emph{after} she creates a folder but \emph{before} she gets a chance to add a file to it, i.e., the file-creation commands are constantly reordered. 

A property one may want to ensure is that at least some commands to eventually \emph{stabilize}, i.e., form an ever growing \emph{stable prefix} of commands that experience only finite number of reorderings for each command.
The stable prefix allows at least some clients to eventually stop having different responses to the same command.
Eventual stability, however,  does not impose restrictions on \emph{how} commands stabilize: the context of a command after the last reordering may be unfavorable for the client. 
In the example above, the stable prefix reorders each file-creation command of Alice, and she effectively \emph{starves}, i.e., none of her \emph{context-sensitive} commands brings the desired effect.  
One might therefore want to enrich stable histories with a notion of \emph{fairness}: intuitively, each client should have at least \emph{some} of its context-sensitive commands never reordered, i.e., stabilize in their \emph{initial contexts}.

In this paper, we aim to account for the previously described effects of \emph{wait-freedom} and to define progress guarantees that can be provided in partition-prone and asynchronous systems---where replicas cannot rely on any sort of coordination (quorum acknowledgment or consensus)---in terms of \emph{stability} and \emph{fairness}.
Our contributions are twofold.  

First, we introduce \emph{eventual state-machine replication} as a formal specification of the problem solved by \emph{wait-free} replicated data types (e.g., CRDTs or optimistic replication algorithms). 
This formalization establishes \emph{wait-freedom} as a necessary condition for systems implementing eventual consistency to remain tolerant to partitions of any size.
We further introduce two progress properties: \emph{stability}, ensuring a growing stable prefix of commands, and \emph{fairness}, ensuring that some context-sensitive commands from each live process stabilize in their initial contexts.

The second contribution is a general framework to achieve eventual state-machine replication for any replicated data type. 
It involves replicas maintaining their state as a DAG (directed acyclic graph) grasping the causal relations of the commands the replica is \say{aware of}.
A \emph{reconciliation function} is used to compute the effect of the DAG commands by ordering them in a local history, using a specific arbitrating strategy.    

We present first a low-cost reconciliation function that ensures stability by respecting causality and deterministically ordering concurrent commands.
We then present a more computationally costly reconciliation function that, additionally, ensures fairness by iteratively selecting certain \say{leader} vertices in the DAG and appending their causal past to the current history. 
Leader vertices are chosen according to a round-robin order of their issuing process to ensure fairness. 
Interestingly, this approach shares some similarities with DAG-Rider~\cite{dagrider} and subsequent DAG-based blockchain protocols, albeit designed for a different, eventually synchronous setting. 

To sum up, this paper:
\begin{itemize}
    \item specifies \emph{eventual state-machine replication} as the problem solved by wait-free replicated data types such as CRDTs or optimistic replication algorithms; 
    \item augments the specification of eventual state-machine replication with \emph{stability and fairness};
    \item presents a \emph{DAG-based framework} that enables eventual state-machine replication;
    \item proposes \emph{asynchronous solutions} to ensure stability and fairness in this framework. 
\end{itemize}

The rest of this paper is organized as follows. 
We discuss related work in Section~\ref{sec:relatedwork}.
The system model we consider and the specification are described respectively in Section \ref{sec:model} and \ref{sec:problem}.
We present our DAG-based framework along the reconciliation functions we propose to ensure stability and fairness in Section~\ref{sec:solutions}.
Section \ref{sec:conclusion} concludes the paper and overviews future work.
Appendix~\ref{sec-a:experiments} presents the experimental analysis of the computational costs incurred by our reconciliation functions. 


\section{Related Work}\label{sec:relatedwork}
The CAP theorem~\cite{cap,cap-lynch} splits distributed systems in two categories based on what they favor: strong consistency or availability.

\subparagraph*{Strongly Consistent Systems.} For strongly consistent systems, Lamport~\cite{paxos} described a partially synchronous fault-tolerant state-machine replication protocol, Castro and Liskov~\cite{pbft} extended it to the Byzantine setting.
DAG-based blockchains (\cite{aleph,dagrider,narwhal,bullshark}, to name a few) have recently gained momentum due to their stable throughput and elegant separation of data dissemination and ordering.  
We employ a similar mechanism: the issued commands are maintained in a DAG using reliable broadcast and ordered using a reconciliation function.
Moreover, similar to committed vertices in DAG-based blockchains, once a vertex gets into a stable prefix, all its causal past is also getting stable, which gives us "stable throughput": it might take a while for a vertex to get stable because of temporary partitions, but once it does, the whole bunch of its predecessors do.

A recent line of work aims to provide \textit{fair ordering consensus} (for blockchains implementing SMR), motivated by the risk of manipulation in transaction ordering (e.g. front-running, sandwich attacks or censorship) which can lead to unfair economic advantages \cite{sok-transaction-reordering}. The general idea is to observe incoming transactions and reach agreement on a fair ordering, where an ordering is considered fair between any pair of transactions if it preserves their relative order of reception as observed by a majority of (correct) processes \cite{transaction-fairness}. A central challenge for these protocols is that, even when all processes are correct, they may not be able to agree on a fair total order due to Condorcet cycles in the reception order. As a result, weaker forms of fairness have been proposed, where transactions potentially involved in a Condorcet cycle are included in the same \textit{batch}, avoiding strict ordering among them \cite{quick-order,themis}. A more relaxed notion of fairness (sometimes referred to as transaction liveness) only guarantees that a transaction is eventually included in the blockchain. This is typically achieved by aggregating transactions from multiple proposers, either by including several proposals in a single block (e.g., Red Belly \cite{redbelly} ) or by rotating leaders (e.g., HotStuff  \cite{hotstuff}, DAG-Rider\cite{dagrider}). Additionally, in DAG-Rider, the leader does not choose transactions to include in blocks, which further reduces the potential for manipulation. 
However, these forms of fairness do not guarantee fairness to clients, as a given client’s transactions might never be executed due to conflicts or system asynchrony \cite{order-fairness}. 

\subparagraph*{Available Systems.}
Available systems relax strong consistency to preserve availability under network partitions: replicas evolve independently and are guaranteed to converge to a common state \emph{eventually}. \emph{Optimistic replication}~\cite{saito-shapiro-or} is a key paradigm for building such systems.  
Its principle is to apply operations immediately upon receipt and reconcile them later.  
The resulting correctness criterion, \emph{eventual consistency}, was first formalized by Saito and Shapiro~\cite{saito-shapiro-or} as the fundamental property of optimistic replication, and later popularized in large-scale cloud storage systems by Vogels~\cite{vogels-ec}.
Eventual consistency has been instantiated in several practical architectures, including Bayou~\cite{bayou}, Dynamo~\cite{dynamo}, Cassandra~\cite{cassandra}, and numerous implementations of  CRDTs~\cite{crdt11}, which guarantee a specific form of eventual consistency known as \emph{Strong Eventual Consistency (SEC)}.  

However, conventional understanding of eventual consistency does not capture the finer-grained progress notions we introduce in this paper, namely \emph{stability} and \emph{fairness}. 
The conventional definition~\cite{vogels-ec},  
``if no new updates are made to the object, eventually all accesses will return the last updated value''\footnote{Verner Vogels, *Eventually Consistent — Revisited*, \url{https://www.allthingsdistributed.com/2008/12/eventually_consistent.html}} captures convergence in \emph{quiescent} periods, when new commands stop arriving, but is limited to storage-like systems and does not account for executions where updates continuously occur.
Consequently, it cannot express progress properties such as \emph{stability} or \emph{fairness}.  

Saito and Shapiro~\cite{saito-shapiro-or} proposed a broader definition of eventual consistency requiring the existence of an ever-growing \emph{stable prefix} (or ``committed prefix'') of issued operations.  
Assuming fault-free systems, it also implicitly provides a weak form of fairness, requiring that ``every issued operation ... will eventually be included in the committed prefix.''  
However, this still-informal definition does not address the problem of \emph{local progress}: the effects of operations issued by a process may always be revoked due to concurrency.  
Our correctness criterion generalizes this view to arbitrary sequential objects, applies to all executions (not only eventually quiescent ones), and introduces a stronger, practically meaningful notion of fairness.  

A related concept, the \emph{eventual stable prefix} property, appears in the blockchain literature~\cite{finalityinblockchains}. %
It is ensured through a specific \emph{fork choice rule} that copes with a potentially unbounded number of Byzantine clients.  
However, that work does not address context-specific fairness.
Our reconciliation functions are applied in consensusless asynchronous settings and generalize the fork choice rule to arbitrary data types beyond cryptocurrencies.

Some works are in the intersection of strongly consistent and available systems, by providing strong consistency guarantees to some operations, while allowing other operations to be wait-free~\cite{zeno}, or by providing a speculative result to the user, which needs to later be confirmed by consensus~\cite{asynccommitsemantic,creek}.
However, none of this work tolerate partitions of any size, and provide fairness guarantees to all clients even in the wait-free part of the system.

In a nutshell, our work is the first to formalize fine-grained progress guarantees of available systems, in terms of \emph{stability} and \emph{fairness} and providing a framework to design  partition-tolerant and  asynchronous replication algorithms. 
The choice of eventual consistency, unlike stronger criteria that require deciding on an ordering, is the key to provide such a strong notion of fairness.
 

\section{Model}\label{sec:model}

\subparagraph*{Replicas, Communication Channels.}
We consider a set $\Pi$ of $n$ \emph{replicas} (or \emph{processes}).
We also assume a set of \emph{clients} that submit inputs to the system and are provided with outputs. 
For simplicity, we assume that clients reside directly on the replicas. 
Every process is assigned a deterministic \emph{algorithm}, a sequential automaton that accepts inputs (application calls and messages from other processes) and produces outputs by applying the automaton's state transition function.      
A \emph{run} (or an \emph{execution}) of an algorithm is a sequence of algorithmic \emph{steps}.
The processes are subject to (benign) \emph{crash} failures\footnote{We discuss Byzantine failures in Section~\ref{sec:conclusion}.}: a faulty process prematurely stops taking steps of its algorithm, simply ignoring the inputs it receives from some point on. 
A process that never fails in a given run is called \emph{correct}.
We make no assumptions on the number of faulty processes.

The processes communicate over \emph{reliable} point-to-point channels~\cite{cachin2011introduction}.   
We make no synchrony assumptions, i.e., the messages between correct processes are eventually received, but there is no bound on the communication delay.

For simplicity of analysis, we specify the properties of the system using a \emph{global clock} that assigns monotonically increasing \emph{times} to the steps in a run, and no process has access to the clock.
Let $x(t)$ denote the value of variable $x$ at time $t$.  

\subparagraph*{Data Types.}
Our goal is to replicate a service across many processes.
A service is formalized as a \emph{data type}, a state-machine $\rdt$, where:
\begin{itemize}
    \item $Q$ is a set of \emph{states}, $q_0\in Q$ is the \emph{initial state};
    \item $O$ is a set of \emph{operations}, $R$ is a set of \emph{responses};
    \item $\sigma: Q\times O \rightarrow Q \times R$ is a \emph{state transition function} that associates each state and operation (applied to it) with the resulting state and the produced response.
\end{itemize}
%
%
%

\subparagraph*{Reliable Broadcast.}
The processes are equipped with a \emph{reliable broadcast}~\cite{cachin2011introduction} primitive that exports a call $r\_broadcast(m)$ and an upcall $r\_deliver(m)$ for each \emph{message} $m$ and satisfies:
\begin{itemize}
    \item \textbf{RB-Integrity}: If a process delivers a message $m$ from $p_s$, then $m$ was previously broadcast by $p_s$ and $m$ is delivered no more than once;
    \item \textbf{RB-Validity}: If $p_s$ is correct, and $p_s$ broadcasts a message $m$, then $p_s$ eventually delivers $m$;
    \item \textbf{RB-Totality}: If a correct process delivers a message $m$, then every correct process eventually delivers $m$.
\end{itemize}
Reliable broadcast can be implemented in an asynchronous system, regardless of the number of faulty processes~\cite{cachin2011introduction}.


\section{Eventual State-Machine Replication}
\label{sec:problem}
%
%
%
%

\subparagraph*{Eventual State-Machine Replication.}
The eventual state-machine replication problem is specified as follows.
Let $\textit{dt} = \rdt$ be any data type, specified as a state-machine, to be replicated.
An algorithm solving \emph{eventual state-machine replication} equips every replica $i \in \Pi$ with a procedure $\textit{append}(o)$, where $o\in \operations$, which issues a command $c$ to the replicated state-machine.
A command $c$ is a tuple $\command$, where $o$ is the operation, $i\in\Pi$ is the replica issuing $c$, and $s$ is a local \emph{sequence number} assigned by $i$ (to distinguish multiple commands invoking the same operation).

An \emph{eventual state-machine replication} algorithm provides each replica $i$ with a local \emph{history} $H_i$ of commands.
The value of $H_i$ at time $t$ is denoted $H_i(t)$ and given $\textit{dt}$, $H_i(t)$ uniquely determines the resulting state and the response of every command in it, at time $t$. 
We denote by $\set{H_i(t)}$ the unordered set of all commands in the sequence $H_i(t)$.
Every \emph{eventual state-machine replication} algorithm run ensures:

\begin{itemize}
    
\item {\textbf{Validity:}} For all correct $i\in \Pi$ and all times $t$, $H_i(t)$ only contains issued commands, i.e., for each $c = (o,j,s)\in H_i(t)$, $o$ is the $s$-th operation issued by $j$ and $c$ is unique.   
%

\item {\textbf{Monotonicity:}} For all correct replicas $i\in\Pi$ and times $t$, $\set{H_i(t)}\subseteq \set{H_i(t+1)}$. 

\item {\textbf{Totality:}} For all correct $i,j\in \Pi$ and times $t$, there exists a time $t'$ such that $\set{H_i(t)} \subseteq \set{H_j(t')}$.

\item {\textbf{Convergence:}} For all correct $i,j\in \Pi$ and times $t, t'$, $\set{H_i(t)}=\set{H_j(t')} \implies H_i(t)=H_j(t')$. 

\item {\textbf{Wait-freedom:}} Let $o$ be the $s$-th operation issued by a correct replica $i$ at time $t$, then $(o,i,s)\in H_i(t)$.

\end{itemize}

Let us note that every run produced by an eventual state-machine replication algorithm trivially satisfies Strong Eventual Consistency (SEC)~\cite{crdt11}\footnote{Totality here corresponds to Eventual delivery in SEC, and Convergence---to Strong Convergence in SEC~\cite{crdt11}.}.
Our specification formally extends SEC with wait-freedom, which was until now an implicit expectation for CRDTs under the name \say{high-availability}.
We present next some consequences of wait-freedom on the progress of the replicated data type.

\subparagraph*{Stable and Fair Progress.}
The specification above provides a very weak form of progress, comparable with progress requirements of reliable broadcast~\cite{cachin2011introduction}.
Every command issued by a correct process eventually gets into the local history of every correct process, but there are no guarantees on the order in which commands are placed.
We focus on infinite runs, where commands can be reordered infinitely often and hinder the progress of the whole replication. 

More in details, let $H_i(t)$ be the local history of process $i$ when it issues a command $c$. 
By wait-freedom, $c$ belongs to $H_i$ as soon as it is issued.
We define the \emph{context} of $c$ at any time $t'$, denoted as $C_c(t')$, the local prefix of $H_i(t')$ up to $c$.
The \emph{initial context} of $c$ is $C_c(t)$ but it may change over time: $c$ or some commands in $C_c(t)$ may later be moved to positions different from those in $H_i(t)$.
We specifically focus here on commands $c$ that the client declares of as \emph{context-sensitive}. 
Intuitively, the client might want to declare context-sensitive commands which only brings the desired \emph{effect} (the response or the resulting state) in specific context.
For example adding a command adding a file to a folder succeeds if the folder is present and fails otherwise, so Alice in our example might want to declare her file-creation commands as context-sensitive.
If $C_c(t') \neq C_c(t)$ then we say that the context of $c$ has been \textit{reordered}. 
It implies that the effect and response of $c$ in $H_i(t')$ may be different from the ones in $H_i(t)$.
In particular, for data types with explicitly defined illegal states, the command may later be disabled. 

Let us recall that it benefits clients to add guarantees of \emph{stability}: eventually, each command is fixed forever in a context to prevent infinite reordering; and of \emph{fairness}: some commands of each correct client stabilize with their initial contexts (when clients issues infinitely many commands) to prevent clients from starving.
Formally, \emph{eventually fair state-machine replication} is achieved when every run of eventual state-machine replication satisfies:

\begin{itemize}

\item {\textbf{Growing Stable Prefix:}} For two sequences $s,s'$, we write $s\preceq s'$ (resp., $s\prec s'$) if $s$ is a prefix (resp., proper prefix) of $s'$.
If the run has infinitely many inputs, then there exists a series $S_1,S_2,\ldots$ of command sequences, such that for all $\ell \in \Nat$, (i)~$S_{\ell}\prec S_{\ell+1}$, and (ii)~for all correct $i\in\Pi$, there exists a time $t$ such that for all $t'\geq t$, $S_{\ell}\preceq H_i(t')$.

The (growing) \emph{stable prefix} $S_1,S_2,\ldots$ converges to the \emph{stable history} $S$: $\forall \ell\; S_{\ell} \prec S$.
    
\item {\textbf{Fairness:}} If a correct process $i$ issues infinitely many context-sensitive commands, then there exists infinitely many context-sensitive commands $c$ of $i$ such that, for all $t'>t$, $C_c(t) = C_c(t')$, where $t$ is the time when $i$ issued $c$.
\end{itemize}

Let us remark that eventually fair state-machine replication is strictly stronger than SEC: growing stable prefix and fairness are enhanced liveness properties that are absent from traditional SEC implementations. 
As a process includes all previously issued commands in the context of the new one (monotonicity and wait-freedom), fairness also implies that each command issued by a correct process is eventually included in $S$.
For simplicity, we do not explicitly require preserving causality in (fair) eventual state-machine replication, while it can easily be implemented in our DAG framework presented in Section~\ref{sec:solutions}.
In fact both the reconciliation functions we are going to describe respect the causal order.
%


\section{DAGs and Reconciliation Functions} \label{sec:solutions}

We now describe a framework enabling eventual state-machine replication for any given data type $\rdt$. 
This framework is based on a directed acyclic graphs (DAG) collaboratively constructed by the replicas.
Intuitively, the DAG keeps track of issued commands and their causality relations.
A \textit{reconciliation function} is then used to totally order the vertices of the DAG and produce the local history. 

\subsection{Construction}\label{subsec:dagbasedconstruction}

\subparagraph*{Overview.}
A directed graph $D$ is a tuple $(V, E)$, where $V \subset \commandSet$ denotes the vertices of $D$ (a set of commands) and $E \subseteq V\times V$ its edges. 
A vertex $v\in V$ is a command $\command$.
A DAG (directed acyclic graph) is a directed graph without cycles. 
Each replica $i$ maintains a local copy of the DAG, denoted $D_i=(V_i,E_i)$.  
In the following, we may drop the subscript when there is no ambiguity.
Let $\dags$ denote the set of finite DAGs. 

When a replica $i$ receives as input an operation $o\in\operations$, it invokes an \textit{append} procedure, increments its sequence number $s$, adds a new vertex $v=\command$ to its local DAG $D_i$, and directed edges from the current \emph{leaf} vertices of $D_i$ to $v$ (or from the root $\epsilon$ if $D_i$ is empty). 
The edges therefore represent the \emph{happened-before} relations across commands.
The command is considered \emph{issued} once the \textit{append} completes.

If there is a path from $v'$ to $v$ in $D=(V,E)$ (denoted by $v' \leadsto_D v$), we say that $v'$ is in the \emph{causal past} of $v$. 
We denote the induced subgraph of $D$ consisting of vertices in the causal past of $v$ (including $v$ itself since $v \leadsto_D v$) by $\past_D(v)=(\{v'\in V | v' \leadsto_D v\},\{(v',v'')\in E | v'' \leadsto v \land v' \leadsto v \}$.
If $v \not\leadsto_D v'$ and $v' \not\leadsto_D v$, we say that the two commands are \emph{concurrent} in $D$.   

An example of a DAG corresponding to a basic Network File System~\cite{CRDTforconcurrentfilesystems} is depicted in Figure~\ref{fig:nfsDAGExample}.
Clients can create and delete directories: $\operations = \{ \textit{mkdir}(\textit{path}, \textit{name}), \textit{rmdir}(\textit{path})\}$.
Removing is enabled only on directories with no children ($\textit{rmdir}$ returns an error if there are some directories whose paths are prefixed by $\textit{path}$) and creating is enabled only on existing paths ($\textit{mkdir}(\textit{path}, \textit{name})$ returns an error if there is an invalid directory composing $\textit{path}$). 
\begin{figure}
    \centerline{
    \tikzset{every picture/.style={line width=0.75pt}} 

\begin{tikzpicture}[x=0.75pt,y=0.75pt,yscale=-0.85,xscale=0.85]

\draw    (106.98,42.3) -- (31.13,107) ;
\draw [shift={(108.5,41)}, rotate = 139.54] [color={rgb, 255:red, 0; green, 0; blue, 0 }  ][line width=0.75]    (6.56,-1.97) .. controls (4.17,-0.84) and (1.99,-0.18) .. (0,0) .. controls (1.99,0.18) and (4.17,0.84) .. (6.56,1.97)   ;
\draw    (107.13,107) -- (31.13,107) ;
\draw [shift={(109.13,107)}, rotate = 180] [color={rgb, 255:red, 0; green, 0; blue, 0 }  ][line width=0.75]    (6.56,-1.97) .. controls (4.17,-0.84) and (1.99,-0.18) .. (0,0) .. controls (1.99,0.18) and (4.17,0.84) .. (6.56,1.97)   ;
\draw    (105.74,171.69) -- (31.13,107) ;
\draw [shift={(107.25,173)}, rotate = 220.93] [color={rgb, 255:red, 0; green, 0; blue, 0 }  ][line width=0.75]    (6.56,-1.97) .. controls (4.17,-0.84) and (1.99,-0.18) .. (0,0) .. controls (1.99,0.18) and (4.17,0.84) .. (6.56,1.97)   ;
\draw    (219.19,42.25) -- (138.38,107) ;
\draw [shift={(220.75,41)}, rotate = 141.3] [color={rgb, 255:red, 0; green, 0; blue, 0 }  ][line width=0.75]    (6.56,-1.97) .. controls (4.17,-0.84) and (1.99,-0.18) .. (0,0) .. controls (1.99,0.18) and (4.17,0.84) .. (6.56,1.97)   ;
\draw    (219.81,105.76) -- (137.75,41) ;
\draw [shift={(221.38,107)}, rotate = 218.28] [color={rgb, 255:red, 0; green, 0; blue, 0 }  ][line width=0.75]    (6.56,-1.97) .. controls (4.17,-0.84) and (1.99,-0.18) .. (0,0) .. controls (1.99,0.18) and (4.17,0.84) .. (6.56,1.97)   ;
\draw    (219.8,108.23) -- (136.5,173) ;
\draw [shift={(221.38,107)}, rotate = 142.13] [color={rgb, 255:red, 0; green, 0; blue, 0 }  ][line width=0.75]    (6.56,-1.97) .. controls (4.17,-0.84) and (1.99,-0.18) .. (0,0) .. controls (1.99,0.18) and (4.17,0.84) .. (6.56,1.97)   ;
\draw    (219.38,107) -- (138.38,107) ;
\draw [shift={(221.38,107)}, rotate = 180] [color={rgb, 255:red, 0; green, 0; blue, 0 }  ][line width=0.75]    (6.56,-1.97) .. controls (4.17,-0.84) and (1.99,-0.18) .. (0,0) .. controls (1.99,0.18) and (4.17,0.84) .. (6.56,1.97)   ;
\draw    (331.94,171.75) -- (250.63,107) ;
\draw [shift={(333.5,173)}, rotate = 218.53] [color={rgb, 255:red, 0; green, 0; blue, 0 }  ][line width=0.75]    (6.56,-1.97) .. controls (4.17,-0.84) and (1.99,-0.18) .. (0,0) .. controls (1.99,0.18) and (4.17,0.84) .. (6.56,1.97)   ;
\draw   (109.13,107) .. controls (109.13,98.92) and (115.67,92.38) .. (123.75,92.38) .. controls (131.83,92.38) and (138.38,98.92) .. (138.38,107) .. controls (138.38,115.08) and (131.83,121.63) .. (123.75,121.63) .. controls (115.67,121.63) and (109.13,115.08) .. (109.13,107) -- cycle ;
\draw   (1.88,107) .. controls (1.88,98.92) and (8.42,92.38) .. (16.5,92.38) .. controls (24.58,92.38) and (31.13,98.92) .. (31.13,107) .. controls (31.13,115.08) and (24.58,121.63) .. (16.5,121.63) .. controls (8.42,121.63) and (1.88,115.08) .. (1.88,107) -- cycle ;
\draw   (108.5,41) .. controls (108.5,32.92) and (115.05,26.38) .. (123.13,26.38) .. controls (131.2,26.38) and (137.75,32.92) .. (137.75,41) .. controls (137.75,49.08) and (131.2,55.63) .. (123.13,55.63) .. controls (115.05,55.63) and (108.5,49.08) .. (108.5,41) -- cycle ;
\draw   (107.25,173) .. controls (107.25,164.92) and (113.8,158.38) .. (121.88,158.38) .. controls (129.95,158.38) and (136.5,164.92) .. (136.5,173) .. controls (136.5,181.08) and (129.95,187.63) .. (121.88,187.63) .. controls (113.8,187.63) and (107.25,181.08) .. (107.25,173) -- cycle ;
\draw   (221.38,107) .. controls (221.38,98.92) and (227.92,92.38) .. (236,92.38) .. controls (244.08,92.38) and (250.63,98.92) .. (250.63,107) .. controls (250.63,115.08) and (244.08,121.63) .. (236,121.63) .. controls (227.92,121.63) and (221.38,115.08) .. (221.38,107) -- cycle ;
\draw   (220.75,41) .. controls (220.75,32.92) and (227.3,26.38) .. (235.38,26.38) .. controls (243.45,26.38) and (250,32.92) .. (250,41) .. controls (250,49.08) and (243.45,55.63) .. (235.38,55.63) .. controls (227.3,55.63) and (220.75,49.08) .. (220.75,41) -- cycle ;
\draw   (333.5,173) .. controls (333.5,164.92) and (340.05,158.38) .. (348.13,158.38) .. controls (356.2,158.38) and (362.75,164.92) .. (362.75,173) .. controls (362.75,181.08) and (356.2,187.63) .. (348.13,187.63) .. controls (340.05,187.63) and (333.5,181.08) .. (333.5,173) -- cycle ;
\draw   (333,41) .. controls (333,32.92) and (339.55,26.38) .. (347.63,26.38) .. controls (355.7,26.38) and (362.25,32.92) .. (362.25,41) .. controls (362.25,49.08) and (355.7,55.63) .. (347.63,55.63) .. controls (339.55,55.63) and (333,49.08) .. (333,41) -- cycle ;
\draw    (218.75,41) -- (137.75,41) ;
\draw [shift={(220.75,41)}, rotate = 180] [color={rgb, 255:red, 0; green, 0; blue, 0 }  ][line width=0.75]    (6.56,-1.97) .. controls (4.17,-0.84) and (1.99,-0.18) .. (0,0) .. controls (1.99,0.18) and (4.17,0.84) .. (6.56,1.97)   ;
\draw    (331,41) -- (250,41) ;
\draw [shift={(333,41)}, rotate = 180] [color={rgb, 255:red, 0; green, 0; blue, 0 }  ][line width=0.75]    (6.56,-1.97) .. controls (4.17,-0.84) and (1.99,-0.18) .. (0,0) .. controls (1.99,0.18) and (4.17,0.84) .. (6.56,1.97)   ;

\draw (12,103) node [anchor=north west][inner sep=0.75pt]  [xscale=0.75,yscale=0.75] [align=left] {$\displaystyle \epsilon $};
\draw (115,36) node [anchor=north west][inner sep=0.75pt]  [xscale=0.75,yscale=0.75] [align=left] {$\displaystyle v_{A}$};
\draw (115,102) node [anchor=north west][inner sep=0.75pt]  [xscale=0.75,yscale=0.75] [align=left] {$\displaystyle v_{B}$};
\draw (114,169) node [anchor=north west][inner sep=0.75pt]  [xscale=0.75,yscale=0.75] [align=left] {$\displaystyle v_{C}$};
\draw (227,36) node [anchor=north west][inner sep=0.75pt]  [color={rgb, 255:red, 229; green, 0; blue, 25 }  ,opacity=1 ,xscale=0.75,yscale=0.75] [align=left] {$\displaystyle \textcolor[rgb]{0.9,0,0.1}{v}\textcolor[rgb]{0.9,0,0.1}{_{D}}$};
\draw (228,103) node [anchor=north west][inner sep=0.75pt]  [color={rgb, 255:red, 255; green, 0; blue, 0 }  ,opacity=1 ,xscale=0.75,yscale=0.75] [align=left] {$\displaystyle \textcolor[rgb]{0.9,0,0.1}{v}\textcolor[rgb]{0.9,0,0.1}{_{E}}$};
\draw (340,36) node [anchor=north west][inner sep=0.75pt]  [xscale=0.75,yscale=0.75] [align=left] {$\displaystyle v_{F}$};
\draw (340,169) node [anchor=north west][inner sep=0.75pt]  [xscale=0.75,yscale=0.75] [align=left] {$\displaystyle v_{G}$};
\draw (68,10) node [anchor=north west][inner sep=0.75pt]  [xscale=0.75,yscale=0.75] [align=left] {$\displaystyle o_{1} :mkdir( '/',\ 'd 1')$};
\draw (68,122) node [anchor=north west][inner sep=0.75pt]  [xscale=0.75,yscale=0.75] [align=left] {$\displaystyle o_{2} :mkdir( '/',\ 'd2')$};
\draw (68,188) node [anchor=north west][inner sep=0.75pt]  [xscale=0.75,yscale=0.75] [align=left] {$\displaystyle o_{3} :mkdir( '/',\ 'd3')$};
\draw (204,77) node [anchor=north west][inner sep=0.75pt]  [xscale=0.75,yscale=0.75] [align=left] {$\displaystyle o_{5} :mkdir( '/d2',\ 'd4')$};
\draw (192,10) node [anchor=north west][inner sep=0.75pt]  [xscale=0.75,yscale=0.75] [align=left] {$\displaystyle o_{4} :rmdir( '/d2')$};
\draw (253,188) node [anchor=north west][inner sep=0.75pt]  [xscale=0.75,yscale=0.75] [align=left] {$\displaystyle o_{7} :mkdir( '/d2/d4',\ 'd5')$};
\draw (298,10) node [anchor=north west][inner sep=0.75pt]  [xscale=0.75,yscale=0.75] [align=left] {$\displaystyle o_{6} :rmdir( '/d1')$};

\end{tikzpicture}}
    \caption{A DAG representing an execution of a basic Network File System. Clients can create (operation $\textit{mkdir}(\textit{path}, \textit{name})$) and delete directories (operation $\textit{rmdir}(\textit{path})$).}
    
    \label{fig:nfsDAGExample}
\end{figure}
Note that in the example in Figure~\ref{fig:nfsDAGExample}, different ordering of the commands in the pair of concurrent vertices $(v_D, v_E)$ (marked in red) produces different outputs. 
Ordering $v_D$ before $v_E$ causes the output of $o_5$ to be an error, and vice versa. 
Such an ordering is determined in our framework using a \emph{reconciliation function} that we define next.

\subparagraph*{Reconciliation Function.} 
The state of the replicated object at time $t$, as well as the response of every command can be computed by applying a \emph{reconciliation function $f$} to $D_i(t)$. 
The function arranges the vertices of $D_i$ into the local history $H_i$. 
We denote by $\commandSet^*$ the set of sequences of commands.
Note that $f$ can implement different arbitration strategies, for instance, the popular \say{remove wins} of CRDTs~\cite{crdt11,CRDTforconcurrentfilesystems, DagWithoutFinality}. 

\begin{definition}[Reconciliation function]
 A deterministic function $f: \dags \mapsto \commandSet^*$, is a reconciliation function if: 
\begin{itemize}
    \item \emph{\textbf{RF-Totality}}:  $\forall D=(V,E)$, $v \in V\;\Leftrightarrow\; v \in f(D)$.
\end{itemize}
\end{definition}

\subparagraph*{Append Operation.}
Algorithm~\ref{alg:append} presents the pseudo-code of replica $i$ for the $\textit{append}(o)$ procedure.
A new command is created and added to DAG in the vertex $v$.
The local history is updated using the reconciliation function, and the DAG is shared with other replicas using reliable broadcast.  

\begin{algorithm}
    \caption{Appending an operation to the DAG}
    \label{alg:append}
    \begin{algorithmic}[1]
        \Procedure{$\textit{append}$}{$o$}
            \State $(V_i, E_i) := D_i$
            \State $\parents := \{ p \in V_i: \not\exists v'\in V_i: (p, v')\in E_i\}$ \Comment{Leaves of $D_i$}
            \State $s++$
            \State $v := \command$
            \State $D_i := (V_i \cup \{ v \}, E_i \cup \{(p, v), \forall p \in \parents \})$\label{code:appendtodag} \Comment{an edge from each parent}
            \State $H_i := f(D_i)$\label{code:computehistory} 
            \State $r\_\textit{broadcast}(\langle v, \parents \rangle)$ \label{code:appendbroadcast}
        \EndProcedure
        \Statex
        \Procedure{$r\_\textit{deliver}$}{$\langle v, \parents \rangle$} \Comment{From $j \neq i$}
            \State Wait until $\forall p \in \parents: p \in V_i$ where $D_i=(V_i,E_i)$
            \State $D_i := (V_i \cup \{ v \}, E_i \cup \{(p,v), \forall p \in \parents \})$\label{code:appenddelivered}
            \Statex\Comment{Append $v$ and its edges to the local DAG}
            \State $H_i := f(D_i)$
        \EndProcedure
    \end{algorithmic}
\end{algorithm}

Once $i$ delivers a message $\langle v, \parents \rangle$, it adds $v$ with its edges to its DAG $D_j$ and update the local history.
If some vertices from $\parents$ are not yet added in $D_j$, this procedure is delayed until all parent vertices are appended. 

\subparagraph*{Correctness of the Framework.}
We show that our DAG framework equipped with any reconciliation function ensures the properties of eventual state-machine replication. 
We start by proving some auxiliary lemmas. 

\begin{lemma}
\label{lemma:past}
Let $v=(o,i,s)$ be a vertex in a local DAG $D$. 
Let $D'$ be the state of $D_i$ at the moment $i$ issued $v$.
Then $\past_D(v)=\past_{D'}(v)$.
\end{lemma}
\begin{proof}[Proof]
Once $i$ adds $v$ to $D_i$ it uses reliable broadcast to disseminate $v$ with its set $P$ of parents in $D_i$ to other replicas.
By RB-Integrity of reliable broadcast, before adding $v$ to its local graph $D$, every replica first waits until every vertex in $P$ is added to $D$.
The replica then adds $v$ with $\{(p,v), p\in P\}$ to $D$.  
Recursively, the causal past of $v$ in any local graph $D$ is identical to its "original" causal past $\past_{D'}(v)$. 
\end{proof}
Lemma~\ref{lemma:past} allows us to omit the subscript $D$ in the definitions of $\past_D(v)$ and $v\leadsto_D v'$, for all (relevant) vertices $v$ and $v'$, from now on we shall write $\past(v)$ and $v\leadsto v'$.

\begin{lemma}
\label{lemma:containment}
For all correct $i,j\in \Pi$ and times $t$, there exists a time $t'$ such that $D_i(t) \subseteq D_j(t')$.
\end{lemma}
\begin{proof}[Proof]
Let $i$ and $j$ be two correct processes and $V_i$ be the set of vertices of $D_i(t)$ at some time $t$.
    By Algorithm~\ref{alg:append}, every vertex $v \in V_i$ has been added to $D_i$ and then disseminated across the system using reliable broadcast (lines~\ref{code:appendtodag} and~\ref{code:appendbroadcast}) or delivered using reliable broadcast (line~\ref{code:appenddelivered}).
    Recursively, every vertex in $\textit{parents}$ is in $V_i$ and, thus, has been earlier $\textit{r\_broadcast}$ or $\textit{r\_delivered}$ by $i$. 
    By Lemma~\ref{lemma:past}, every process that adds $v$ to its local DAG, also adds the same set of edges.  
    Thus, by the RB-Validity and RB-Totality properties of reliable broadcast, at some time $t'$, $j$ must contain all vertices and edges present in $D_i(t)$, i.e., $D_i(t) \subseteq D_j(t')$.
\end{proof}

\begin{lemma}
\label{lemma:progress}
Let $j$ be a process that issues infinitely many commands.
For all times $t$ and correct processes $i$, there is a time $t'> t$ and a vertex $v$ issued by $j$ in $D_j(t')$ such that $D_i(t)\subseteq \past(v)$.
\end{lemma}
\begin{proof}[Proof]
Let $i$ be a correct process and $j$ be a correct process that issues infinitely many commands. 
By Lemma~\ref{lemma:containment}, for all times $t$, there exists $t''$ such that $D_i(t)\subseteq D_j(t'')$. 
Thus, every command issued by $j$ after time $t''$ will result in a vertex $v$ added to $D_j(t')$, $t'\geq t''$, such that $D_i(t)\subseteq D_j(t'') \subseteq \past(v)$. 
\end{proof}

\begin{theorem}\label{thm:reconciliationfunctions}
For any reconciliation function, Algorithm~\ref{alg:append} implements eventual state-machine replication.
\end{theorem}
\begin{proof}[Proof] Consider any execution of Algorithm~\ref{alg:append} equipped with a reconciliation function $f$.
\begin{itemize}
    \item Validity: Every DAG vertex is labelled with a distinct command issued via Algorithm~\ref{alg:append}.
    
    \item Monotonicity: The claim follows from the fact that we can only add new vertices to a DAG (Algorithm~\ref{alg:append}, line~\ref{code:appendtodag}) and the RF-Totality property of $f$ (line~\ref{code:computehistory}).
    
    \item Totality: 
    Let $i$ and $j$ be correct processes.
    By Lemma~\ref{lemma:containment}, for each time $t$, there exists a time $t'$ such that $D_i(t) \subseteq D_j(t')$.
    By RF-Totality of $f$, we have $\set{H_i(t)} \subseteq \set{H_j(t')}$.
    
    \item Convergence: RF-Totality and Lemma~\ref{lemma:past} imply that if $\set{H_i(t)}=\set{H_j(t')}$ then $D_i(t) = D_j(t')$. %
    The claim follows from $f$ being the same function at $i$ and $j$.

    \item Wait-freedom: Let $t$ be the time at which $i$ adds the vertex $v$ produced by its call to Algorithm~\ref{alg:append} to $D_i(t)$. Thus, $(o, i, s) \in H_i(t)$ by RF-Totality of $f$ at line~\ref{code:computehistory}.
\end{itemize}

Thus, Algorithm~\ref{alg:append} satisfies all the properties of eventual state-machine replication.
\end{proof}

\subsection{Stability and Fairness}\label{sec:functions}
We present here two reconciliation functions, \fbfs~(Algorithm~\ref{alg:functionbfs}) ensuring growing stable prefix and \ffair~(Algorithm~\ref{alg:functionfair}) ensuring fairness. 
While \ffair also ensure growing stable prefix, the interest of \fbfs is its lower computational cost. 
Indeed, \fbfs only explores a DAG once from the root, while \ffair does multiple explorations from different vertices.

\subparagraph*{Stable Reconciliation Function \fbfs.}\label{sec:fbfs}
Even though it appears natural, growing stable prefix is not entirely trivial to achieve, as a reconciliation function might choose to prioritize the last issued command.
But, as we show in Algorithm~\ref{alg:functionbfs}, growing stable prefix can be achieved once we establish an order based on the \emph{greatest distance} from the root $\epsilon$ to each vertex $v$, denoted by $\dist(D,v)$.\footnote{For a DAG $D=(V,E)$, the function is computed recursively as: $\dist(D,\epsilon)=0$;  $\forall v\neq\epsilon$, $\dist(D,v)=\max(\{\dist(D,v') | (v',v)\in E\})+1$.} 
Notice that the partial order generated by these distances generalizes the causal order and iterating over
this partial order results in a Breadth First Search (BFS). 
We iteratively go over the vertices at the same distances from the root, from closer to farther ones, each time extending the resulting history $\seq$ with a new command $v$ ($\lambda$ denotes the empty sequence).
Vertices at the same distance are processed based on the identifiers of their issuers: $v \nearrow_{id} v'$ ($v$ precedes $v'$) if  $v = (op,id,sn) \land v' = (op',id',sn') \land id < id'$.

\begin{algorithm}
    \caption{Distance-based reconciliation function \fbfs}
    \label{alg:functionbfs}
    \begin{algorithmic}[1]
        \Procedure{\fbfs}{DAG $D = (V, E)$}
        \State $\seq := \lambda$
        \State $\length := \max(\{ \dist(D,v): v\in V\})$\label{code:sortbydistancestart}
        \For{$d = 1,\ldots,\length$} \label{code:iterationBFS}
            \State $\concurrent := \{ v \in V: dist(D, v) == d \}$ \label{code:concurrentset}
            \While{$\concurrent \not= \emptyset$}
                \State $v := \text{smallest in }\concurrent \text{ by } \nearrow_{id}$ \label{code:concurrentsorting}
                \State $\concurrent := \concurrent \setminus \{ v \}$
                \State $\seq := \seq.v$ \label{code:fbfsfinalappend}
            \EndWhile
            \label{code:concurend}
        \EndFor
        \State\Return $\seq$
        \EndProcedure
    \end{algorithmic}
\end{algorithm}

\begin{lemma}[Bounded same distance set]\label{lemma:boundedsamedistance}
At any time $t$ and at any correct process $i$, $D_i(t) = (V, E)$ satisfies $\forall k \in \Nat, |\{ v \in  V: \dist(D,v) = k\}| \leq n$ where $n = |\Pi|$.
\end{lemma}
\begin{proof}[Proof]
Processes respect their program order: any two commands issued by a same process are causally related and, therefore, cannot be at the same distance from the root.
\end{proof}

\begin{lemma}[Same distance stability]\label{lemma:samedistancestability}
For any $k \in \Nat$, there is a time $t$ such that for every correct process $i$, $D_i(t)=(V,E)$ and $\forall$ time $t' > t$, $D_i(t')=(V',E')$ such that $\{ v \in V: \dist(D_i(t),v) = k\} = \{ v \in V': \dist(D_i(t'),v) = k \}$. 
\end{lemma}
\begin{proof}[Proof]
Let us first observe that for all time $t<t'$ and correct process $i$, $D_i(t)=(V,E)$ and $D_i(t')=(V',E')$ satisfy: $\forall k \in \Nat, \{ v \in V: \dist(D_i(t),v) = k\} \subseteq \{ v \in V': \dist(D_i(t'),v) = k \}$ because $V \subseteq V'$ by Algorithm~\ref{alg:append}.
The observation follows from Lemma~\ref{lemma:past} since $\dist(D,v)$ depends solely on $\past_D(v)$.
The claim is then implied by Lemmas~\ref{lemma:progress}~and~\ref{lemma:boundedsamedistance}.
\end{proof}

\begin{theorem}[\fbfs ensures stability]\label{thm:fbfs}
Every execution of Algorithm~\ref{alg:append} with \fbfs satisfies growing stable prefix.
\end{theorem}
\begin{proof}[Proof]
For a correct process $i$, Lemma~\ref{lemma:samedistancestability} implies that $\forall d \in \Nat$ there is a time $t$ such that $\forall$ time $t' \geq t, \forall d' \leq d$, $concurrent$ for distance $d'$ is fixed (line~\ref{code:concurrentset}).
Because $\nearrow_{id}$ is deterministic, it produces the same order for the same set of vertices at distance $d'$ (line~\ref{code:concurrentsorting}).
Thus, the procedure gives a growing stable prefix $S_1,S_2,\ldots$.
\end{proof}

Clearly, \fbfs is not fair, as it favors commands issued by processes with smaller ids, and it may happen that processes with larger ids are always reordered.
Note that fairness cannot be guaranteed with any order on commands at the same distance, e.g., ordering by the type of operation invoked (\say{remove-wins} arbitration~\cite{crdt11}) or permuting the natural order on ids, i.e., where the \say{smaller} id is the distance from the root modulo the number of processes.
Indeed, one process may always issue commands with an underprivileged type of operation, or always include in its context a command from another process with \say{large} id at distance $d-1$, even though the process has the \say{smallest} id at distance $d$.

\subparagraph*{Fair Reconciliation Function \ffair.}
The reconciliation function described in Algorithm~\ref{alg:functionfair} iteratively constructs a history $\seq$, starting from the empty sequence $\lambda$, as follows.
In each iteration, we locate the next process in the round-robin order with a context-sensitive vertex that causally succeeds every command in $\seq$.
If such a process $j$ exists, we pick the earliest such vertex $v'$ and extend $\seq$ with $v'$ and all \emph{new} vertices in the causal past of $v'$ (the vertices in $\past(v')\setminus\set{\seq}$), ordered in some deterministic way that preserves the causal partial order induced by the edges of $D$.  
We denote this ordering function by $\textit{sort}$.\footnote{As $D$ is acyclic, such a topological sorting exists.} 
If there is no such a process, we extend $\seq$ with the ordered sequence of all remaining vertices, i.e., $V\setminus \set{\seq}$.   

By Lemma~\ref{lemma:progress}, as $D$ eventually contains each command issued by a correct process, for any such fixed $\seq$, every process that issues sufficiently many operations will eventually have a vertex in $D$ that causally succeeds $\seq$.
We show below that this implies stability and fairness. 

Let us come back to the NFS example described in the introduction. 
If Alice declares her file-creation commands context-sensitive, then our algorithm guarantees that eventually she will manage to add her file to a new folder, even if she is slow and unlucky with scheduling.

\begin{algorithm}[t]
\caption{Fair reconciliation function \ffair}
        \label{alg:functionfair}
    \begin{algorithmic}[1]

    \Procedure{\ffair}{DAG $D = (V, E)$}
        \State $seq := \lambda$ \Comment{empty sequence}
        \While{there is a process $i$ satisfying $P$: $i$ issued a vertex $v=(o,i,s)\in V$ such that $o$ is context-sensitive and $\set{\seq}\subseteq \past(v)$}
            \State select the next process $j$ satisfying $P$ in round robin
            \State let $v'$ be the \emph{closest} such vertex of $j$ (in distance to the last vertex of $\textit{seq}$)
            \State $\textit{update} := \textit{sort}(\past(v')\setminus\set{\seq})$ 
            \State $\seq := \seq.\textit{update}$ \Comment{$\seq$ extended}  \label{code:newappended}
        \EndWhile    
        \State $\textit{update} := \textit{sort}(V \setminus \set{\seq})$
        \State $\seq := \seq.\textit{update}$  \label{code:remainingappended}     
        \State\Return $\seq$
    \EndProcedure
    \end{algorithmic}
\end{algorithm}

\begin{theorem}[\ffair ensures stability and fairness]\label{th:ffairstable}
Algorithm~\ref{alg:append} with \ffair ensures growing stable prefix and fairness.
\end{theorem}
\begin{proof}[Proof]
Consider any execution of Algorithm~\ref{alg:append} with \ffair as a reconciliation function.
Let $\tilde D$ denote the (possibly infinite) \emph{limit} DAG to which local DAGs $D_i$ maintained by the correct processes $i$ converge: for all times $t$ and correct processes $i$, $D_i(t) \subseteq \tilde D$.
By Lemma~\ref{lemma:containment}, such a limit DAG exists.
Let $i$ be a process that issues infinitely many commands. 
Let $S_{\ell}$ denote the value of $\seq$ after the $\ell$-th iteration of Algorithm~\ref{alg:functionfair} (line~\ref{code:newappended}) applied to the (infinite) limit DAG $\tilde D$.
By Lemma~\ref{lemma:progress}, every correct process (and $i$ in particular) has a vertex in $\tilde D$ that causally succeeds every vertex in $S_{\ell}$.   
Thus, the construction produces longer and longer histories: $\forall \ell$, $S_{\ell}\prec S_{\ell+1}$ for all $\ell$.
Let $v_{\ell}$ denote the last command in $S_{\ell}$. 
By construction, for all $\ell$, $v_{\ell}\leadsto v_{\ell+1}$.

Suppose, without loss of generality, that it is up to $i$ to add a vertex to $\seq$ in iteration $\ell+1$ (we just wait until it is $i$-th turn in the round-robin order).
Thus, a command $v=v_{\ell}$ of $i$ will end $S_{\ell+1}$.
Moreover, $S_{\ell+1}$ is a topological sorting of $\past(v)$ where $v$ is the last vertex.    
By Lemma~\ref{lemma:past}, $S_{\ell}$ is precisely $H_i(t)$, where $t$ is the time when $i$ issued $v$.    
Thus, the effect of $v$ in $S_{\ell}$ is the effect of $v$ witnessed by $i$ when it issued the command.

By Lemma~\ref{lemma:progress}, $\forall \ell$, every correct process $i$ eventually gets vertex $v_{\ell}$ in its DAG.
Once this happens, every history constructed by $i$ will be an extension of $S_{\ell}$---hence the property of growing stable prefix. 

By the arguments above, every correct process $i$ obtains infinitely many vertices in $v_1,v_2,\ldots$.
Thus, the growing stable prefix will give infinitely many commands never reordered to $i$---hence the property of fairness.
\end{proof}

\subsection{Complexity Analysis}

\subparagraph*{Local Computation.}
In our framework, as in conventional CRDT implementations~\cite{pureopbased}, a reconciliation function is applied to the current DAG each time the state needs to be computed.
%
As DAGs may grow without bound, the local computation can thus incur a considerable cost and make the system impractical.

The reconciliation function \fbfs iterates over sets of vertices at the same distance from the root. 
To avoid exploring the DAG each time \fbfs is computed, we assume processes store in memory the distance of each vertex to the root and the length of the DAG, which they update only when adding a new vertex with a leaf as parent.
Thus, processes only need to compute the ordering $\nearrow_{id}$ for commands at the same distance.
Using merge sort, this can be done in $O(n\log n)$, where $n = |\Pi|$ is the number of processes because there is at most $n$ vertices at the same distance from the root (Lemma~\ref{lemma:boundedsamedistance}, one vertex by process).
The computational complexity of \fbfs is thus $O(kn\log n)$, where $k \leq |V|$ is the length of the DAG and $V$ the set of vertices of the DAG.

In practice, one may chose to optimize \fbfs by computing the ordering of vertices at the same distance from the root only when a new vertex is added to the DAG, and to store in memory these sorted sets of vertices.
The $O(n\log n)$ computational cost would be paid when adding a new vertex to the DAG, such that computing \fbfs comes down to concatenating all the sorted sets, which, using a linked list, amounts to $O(k)$ computational steps.

Adding fairness comes with a cost: \ffair explores the DAG from different sources to find vertices from specific processes in round robin---\say{leader} vertices.
\ffair iteratively selects leader vertices, and for each leader vertex $v$, we need to explore $v$'s entire causal future to choose the next leader vertex $v'$ because the next vertex including $v$ in its causal past may be in a leaf.
The worst case is as follows: the DAG is a path from the root to a single leaf and for every leader vertex $v$, the closest vertex $v'$ from the next starving process is the child of $v$.
In this case, \ffair would select $|V|$ leader vertices and explore the whole path starting from each leader vertex $v$. 
Each exploration requires one less steps than the previous one, thus computing \ffair requires $O(|V|(|V|+1)/2) = O(|V|^2)$ computational steps.

These local complexities allow to guarantee stability and fairness in eventual state-machine replication.
As it is a new concept, our goal was to show what guarantees one may expect and proving lower bounds on the local complexities required for each property is left for future work.

\subparagraph*{Checkpointing.}\label{sec:commitment} Determining that a history prefix gets stable enables stronger consistency guarantees, such as \emph{checkpointing}, allowing the \say{garbage-collection} of the stable sub-DAG and the computation of the reconciliation functions on the \say{unstable} part only.
In the asynchronous model, however, we cannot always ensure that replicas eventually learn when a non-empty prefix is stable without consensus\footnote{Semantic compaction and causal stability~\cite{pureopbased} are CRDTs garbage-collection techniques but can be used only if, respectively, the semantic of the data type is restricted (admits an obsolescence relation), and the execution is fault-free. We focus on generic data types and fault-prone systems.}, even if we ensure that partitions are excluded (by, e.g., assuming that a majority of processes are correct)~\cite{ec-wfd,zeno}.
Indeed, detecting when a prefix gets stable implies a consensus algorithm: to propose its value $v$, a process issues a single command containing $v$; and processes decide the value in the first position of the prefix, once its stability is known.
This reduction does not contradict the guarantee that a prefix stabilizes without consensus, as long as the processes are \emph{unaware of what this stable prefix is}---as achieved in our algorithms (Section~\ref{sec:functions}), as well as in earlier work~\cite{finalityinblockchains, unconsciouseventualconsistency}.

Formally, this precludes asynchronous checkpointing of the system state and forces us to apply the reconciliation function to the entire (ever-growing) DAG. 
Under extra assumptions, we can, however, enrich the system with a \emph{commitment} mechanism that periodically marks longer and longer stable prefixes as \emph{committed} that can be used for computing checkpointing states and truncating the DAGs used for reconciliation.
One way to achieve this is to run an optimistic consensus algorithm in parallel with our DAG-based framework, like~\cite{asynccommitsemantic}: the local history computed by the reconciliation function is divided in a committed prefix and an unstable suffix.
Replicas execute the wait-free framework described in this paper, but also propose in parallel their changes in the suffix to a series of consensus instances.
A sequence in the suffix decided by consensus is moved to the committed prefix.
This requires to adopt the assumptions needed by the consensus algorithm running in parallel, e.g., the existence of the \emph{eventual failure detector}~\cite{ec-wfd} or partial synchrony~\cite{zeno,creek}, provided that a majority of processes are correct~\cite{cap,cap-lynch}. 

This technique achieve commitment of stable prefixes, but preserving fairness in the committed prefix is an open question: 
a starving replica may always issue its context-sensitive commands after a prefix is committed on a different context.

\section{Concluding Remarks}\label{sec:conclusion}
In this paper, we specified eventual state-machine replication and added guarantees of stability (replicas share a growing stable prefix of commands) and fairness (some commands issued by all correct replicas stabilize within their initial contexts).
We described a DAG-based framework where any reconciliation function allows to implement eventual state-machine replication and proposed a function satisfying stability and another function satisfying stability and fairness without relying on consensus or quorums.
We provide a modest empirical study of our framework in Appendix~\ref{sec-a:experiments} which exhibits that, compared to one satisfying only stability, a reconciliation function could reorder a little bit more in order to satisfy fairness.
In executions with partitions, guaranteeing fairness improved the number of commands that could be applied without changing their effects and results.
Some interesting questions are left for future work. 
As discussed in Section~\ref{sec:commitment}, a commitment mechanism should be considered in order to make the system practical, but preserving fairness in the committed prefix an open question.

\subparagraph*{Conflicts and Contexts.}
Our framework opens a lot of space for optimizations.
Performance is heavily affected by the amount of reordering, and it can be considerably improved by reducing the number of context-sensitive commands that require to be treated fairly. 
A natural extension of this work is to design reconciliation functions that can account for concurrent commands and enforce ordering only when \textit{conflicts} occur.
Indeed, there is no reason to execute concurrent \emph{commuting} commands in total order. 
Also, the very notion of a context may be operation-specific, an operation may reach the desired effect even if its causal past is modified, as long as it meets some predicate (e.g., the folder recently created by Alice is still there).
Instead of a growing stable prefix, we may then build a growing stable \emph{equivalence class} of sequences of commands~\cite{lamport2005generalized,moraru13epaxos}. 

\subparagraph*{Byzantine Nodes.}
We expect that our approach can be naturally extended to encompass Byzantine failures, similarly to~\cite{makingCRDTBFT}.
Indeed, the advantage of our consistency guarantees to only be eventual mitigates the notorious \emph{equivocation} problem inherent to strongly consistent systems.  
As typical in this setting, assume that correct nodes digitally sign messages and relay received messages to each other: each time a Byzantine node issues identical commands with different causal \textit{past}, it will eventually be detected by the correct nodes, and the vertices of this node may be removed from their DAGs.\footnote{Note that we do have to rely upon (strongly consistent) Byzantine reliable broadcast~\cite{br85acb} that assumes more than $2/3$ of the nodes to be correct.}  

Assuming only finitely many Byzantine nodes, our fair reconciliation function can be then applied to the resulting DAGs, which would ensure that each correct node gets infinitely many of her commands stabilized in their initial contexts.
Note that this would also mitigate the attempts of a malicious node to deliberately issue commands that trigger reordering to her advantage~\cite{DagWithoutFinality,sandwich}.

\bibliographystyle{plainurl}
\bibliography{refs}

\appendix

\section{Experimental Evaluation}\label{sec-a:experiments}
In this section, we experimentally evaluate the reconciliation functions \fbfs and \ffair.
We first study their computational overhead using two metrics: the number of reorderings they generate in the local history (each reordering may induce a recomputation of the state), and the cost of the reconciliation function itself as the DAG grows. 
We then assess their fairness under different network configurations and varying numbers of processes. 
In appendix~\ref{sec-a:allcontextsensitive}, we first evaluate the reconciliation functions independently of any specific data type, i.e., assuming arbitrary operations and that commands are all context-sensitive.
In appendix~\ref{sec-a:set}, we repeat similar experiments on a concrete data type, namely a set, in order to assess the impact of the semantics of a widely used data type. 

\subparagraph*{Experimental Setup.}
The framework, implemented in Rust,  
was experimentally evaluated 
on a machine with OS Ubuntu 24.04.3, CPU Intel Core Ultra 7 165H (16 cores, including 2 running at 2.5 GHz, 8 at 3.8 GHz, 6 at 5 GHz; and 22 threads) and 32 GB of RAM.
We virtualized processes with Docker~\cite{docker}, using one container per process.
Asynchrony was enforced by processes issuing commands at random intervals and messages were delayed following a normal distribution at the operating system level by Linux's utility \emph{tc (traffic control)}\footnote{\url{https://man7.org/linux/man-pages/man8/tc.8.html}}.
Since our main issue is not scalability, we believe $16$ processes are enough to study fairness.
The necessary material to reproduce the experiments is available online\footnote{On \href{https://github.com/maxper4/fair-wait-free-replicated-data-types-experiments/tree/docker-experiments}{github}, archived at \href{https://archive.softwareheritage.org/swh:1:dir:7f5598679623666993670564cc449edd0b4f97c0;origin=https://github.com/maxper4/fair-wait-free-replicated-data-types-experiments;visit=swh:1:snp:eb182b0c08205dab39b3b05d721e47cd2addf120;anchor=swh:1:rev:e1257a4b09ac88a9bb350844995e1afde3386d0e}{swh:1:dir:7f5598679623666993670564cc449edd0b4f97c0}}.

\subsection{Only Context-sensitive Commands}~\label{sec-a:allcontextsensitive}

\subparagraph*{Reordering and Cost of Reconciliation.} In Figure~\ref{fig:stabilityvsprocessesnosem}, we plot the average number of reorderings per command for \fbfs and \ffair, varying the number of processes. 
We observed that \fbfs generates fewer reorderings than \ffair and thus exhibits higher stability in a quantitative sense. 
This is because enforcing fairness in \ffair requires prioritizing commands issued by slow starving processes, even when these commands are received late, which might incur additional reorderings. 
We also observed that, due to more concurrent commands, increasing the number of processes induces more reorderings.

\begin{figure}
    \centerline{
\setlength{\unitlength}{0.120450pt}
\ifx\plotpoint\undefined\newsavebox{\plotpoint}\fi
\ifx\transparent\undefined%
    \providecommand{\gpopaque}{}%
    \providecommand{\gptransparent}[2]{\color{.!#2}}%
\else%
    \providecommand{\gpopaque}{\transparent{1.0}}%
    \providecommand{\gptransparent}[2]{\transparent{#1}}%
\fi%
\begin{picture}(2159,960)(0,0)
\miterjoin\buttcap
\color[rgb]{0.50,0.50,0.50}
\sbox{\plotpoint}{\rule[-0.200pt]{0.400pt}{0.400pt}}%
\linethickness{0.4pt}%
\multiput(379,241)(24.907,0.000){68}{\usebox{\plotpoint}}
\put(2059,241){\usebox{\plotpoint}}
\sbox{\plotpoint}{\rule[-0.600pt]{1.200pt}{1.200pt}}%
\linethickness{1.2pt}%
\Line(379,241)(420,241)
\Line(2059,241)(2018,241)
\put(346,241){\makebox(0,0)[r]{$0.125$}}
\sbox{\plotpoint}{\rule[-0.200pt]{0.400pt}{0.400pt}}%
\linethickness{0.4pt}%
\multiput(379,334)(24.907,0.000){68}{\usebox{\plotpoint}}
\put(2059,334){\usebox{\plotpoint}}
\sbox{\plotpoint}{\rule[-0.600pt]{1.200pt}{1.200pt}}%
\linethickness{1.2pt}%
\Line(379,334)(420,334)
\Line(2059,334)(2018,334)
\put(346,334){\makebox(0,0)[r]{$0.25$}}
\sbox{\plotpoint}{\rule[-0.200pt]{0.400pt}{0.400pt}}%
\linethickness{0.4pt}%
\multiput(379,427)(24.907,0.000){68}{\usebox{\plotpoint}}
\put(2059,427){\usebox{\plotpoint}}
\sbox{\plotpoint}{\rule[-0.600pt]{1.200pt}{1.200pt}}%
\linethickness{1.2pt}%
\Line(379,427)(420,427)
\Line(2059,427)(2018,427)
\put(346,427){\makebox(0,0)[r]{$0.5$}}
\sbox{\plotpoint}{\rule[-0.200pt]{0.400pt}{0.400pt}}%
\linethickness{0.4pt}%
\multiput(379,520)(24.907,0.000){68}{\usebox{\plotpoint}}
\put(2059,520){\usebox{\plotpoint}}
\sbox{\plotpoint}{\rule[-0.600pt]{1.200pt}{1.200pt}}%
\linethickness{1.2pt}%
\Line(379,520)(420,520)
\Line(2059,520)(2018,520)
\put(346,520){\makebox(0,0)[r]{$1$}}
\sbox{\plotpoint}{\rule[-0.200pt]{0.400pt}{0.400pt}}%
\linethickness{0.4pt}%
\multiput(379,614)(24.907,0.000){68}{\usebox{\plotpoint}}
\put(2059,614){\usebox{\plotpoint}}
\sbox{\plotpoint}{\rule[-0.600pt]{1.200pt}{1.200pt}}%
\linethickness{1.2pt}%
\Line(379,614)(420,614)
\Line(2059,614)(2018,614)
\put(346,614){\makebox(0,0)[r]{$2$}}
\sbox{\plotpoint}{\rule[-0.200pt]{0.400pt}{0.400pt}}%
\linethickness{0.4pt}%
\multiput(379,707)(24.907,0.000){68}{\usebox{\plotpoint}}
\put(2059,707){\usebox{\plotpoint}}
\sbox{\plotpoint}{\rule[-0.600pt]{1.200pt}{1.200pt}}%
\linethickness{1.2pt}%
\Line(379,707)(420,707)
\Line(2059,707)(2018,707)
\put(346,707){\makebox(0,0)[r]{$4$}}
\sbox{\plotpoint}{\rule[-0.200pt]{0.400pt}{0.400pt}}%
\linethickness{0.4pt}%
\multiput(379,800)(24.907,0.000){2}{\usebox{\plotpoint}}
\put(412,800){\usebox{\plotpoint}}
\multiput(1418,800)(24.907,0.000){26}{\usebox{\plotpoint}}
\put(2059,800){\usebox{\plotpoint}}
\sbox{\plotpoint}{\rule[-0.600pt]{1.200pt}{1.200pt}}%
\linethickness{1.2pt}%
\Line(379,800)(420,800)
\Line(2059,800)(2018,800)
\put(346,800){\makebox(0,0)[r]{$8$}}
\sbox{\plotpoint}{\rule[-0.200pt]{0.400pt}{0.400pt}}%
\linethickness{0.4pt}%
\multiput(379,893)(24.907,0.000){68}{\usebox{\plotpoint}}
\put(2059,893){\usebox{\plotpoint}}
\sbox{\plotpoint}{\rule[-0.600pt]{1.200pt}{1.200pt}}%
\linethickness{1.2pt}%
\Line(379,893)(420,893)
\Line(2059,893)(2018,893)
\put(346,893){\makebox(0,0)[r]{$16$}}
\sbox{\plotpoint}{\rule[-0.200pt]{0.400pt}{0.400pt}}%
\linethickness{0.4pt}%
\multiput(715,211)(0.000,24.907){24}{\usebox{\plotpoint}}
\put(715,786){\usebox{\plotpoint}}
\multiput(715,852)(0.000,24.907){2}{\usebox{\plotpoint}}
\put(715,893){\usebox{\plotpoint}}
\sbox{\plotpoint}{\rule[-0.600pt]{1.200pt}{1.200pt}}%
\linethickness{1.2pt}%
\Line(715,211)(715,252)
\Line(715,893)(715,852)
\put(715,145){\makebox(0,0){4}}
\sbox{\plotpoint}{\rule[-0.200pt]{0.400pt}{0.400pt}}%
\linethickness{0.4pt}%
\multiput(1051,211)(0.000,24.907){24}{\usebox{\plotpoint}}
\put(1051,786){\usebox{\plotpoint}}
\multiput(1051,852)(0.000,24.907){2}{\usebox{\plotpoint}}
\put(1051,893){\usebox{\plotpoint}}
\sbox{\plotpoint}{\rule[-0.600pt]{1.200pt}{1.200pt}}%
\linethickness{1.2pt}%
\Line(1051,211)(1051,252)
\Line(1051,893)(1051,852)
\put(1051,145){\makebox(0,0){8}}
\sbox{\plotpoint}{\rule[-0.200pt]{0.400pt}{0.400pt}}%
\linethickness{0.4pt}%
\multiput(1387,211)(0.000,24.907){24}{\usebox{\plotpoint}}
\put(1387,786){\usebox{\plotpoint}}
\multiput(1387,852)(0.000,24.907){2}{\usebox{\plotpoint}}
\put(1387,893){\usebox{\plotpoint}}
\sbox{\plotpoint}{\rule[-0.600pt]{1.200pt}{1.200pt}}%
\linethickness{1.2pt}%
\Line(1387,211)(1387,252)
\Line(1387,893)(1387,852)
\put(1387,145){\makebox(0,0){12}}
\sbox{\plotpoint}{\rule[-0.200pt]{0.400pt}{0.400pt}}%
\linethickness{0.4pt}%
\multiput(1723,211)(0.000,24.907){28}{\usebox{\plotpoint}}
\put(1723,893){\usebox{\plotpoint}}
\sbox{\plotpoint}{\rule[-0.600pt]{1.200pt}{1.200pt}}%
\linethickness{1.2pt}%
\Line(1723,211)(1723,252)
\Line(1723,893)(1723,852)
\put(1723,145){\makebox(0,0){16}}
\polygon(379,893)(379,211)(2059,211)(2059,893)
\color{black}
\put(676,819){\makebox(0,0)[r]{\fbfs}}
\color[rgb]{0.22,0.49,0.72}
\polygon*(709,803)(882,803)(882,836)(709,836)
\color{black}
\sbox{\plotpoint}{\rule[-0.200pt]{0.400pt}{0.400pt}}%
\linethickness{0.4pt}%
\Line(709,803)(882,803)
\Line(882,803)(882,835)
\Line(882,835)(709,835)
\Line(709,835)(709,803)
\color[rgb]{0.22,0.49,0.72}
\polygon*(635,211)(712,211)(712,265)(635,265)
\color{black}
\polygon(635,211)(635,264)(711,264)(711,211)
\color[rgb]{0.22,0.49,0.72}
\polygon*(971,211)(1048,211)(1048,416)(971,416)
\color{black}
\polygon(971,211)(971,415)(1047,415)(1047,211)
\color[rgb]{0.22,0.49,0.72}
\polygon*(1307,211)(1384,211)(1384,475)(1307,475)
\color{black}
\polygon(1307,211)(1307,474)(1383,474)(1383,211)
\color[rgb]{0.22,0.49,0.72}
\polygon*(1643,211)(1720,211)(1720,733)(1643,733)
\color{black}
\polygon(1643,211)(1643,732)(1719,732)(1719,211)
\put(1179,819){\makebox(0,0)[r]{\ffair}}
\color[rgb]{0.30,0.69,0.29}
\polygon*(1212,803)(1385,803)(1385,836)(1212,836)
\color{black}
\Line(1212,803)(1385,803)
\Line(1385,803)(1385,835)
\Line(1385,835)(1212,835)
\Line(1212,835)(1212,803)
\color[rgb]{0.30,0.69,0.29}
\polygon*(719,211)(796,211)(796,386)(719,386)
\color{black}
\polygon(719,211)(719,385)(795,385)(795,211)
\color[rgb]{0.30,0.69,0.29}
\polygon*(1055,211)(1132,211)(1132,532)(1055,532)
\color{black}
\polygon(1055,211)(1055,531)(1131,531)(1131,211)
\color[rgb]{0.30,0.69,0.29}
\polygon*(1391,211)(1468,211)(1468,601)(1391,601)
\color{black}
\polygon(1391,211)(1391,600)(1467,600)(1467,211)
\color[rgb]{0.30,0.69,0.29}
\polygon*(1727,211)(1804,211)(1804,894)(1727,894)
\color{black}
\polygon(1727,211)(1727,893)(1803,893)(1803,211)
\put(24,552){\rotatebox{-270}{\makebox(0,0){Average number of reordering}}}
\put(90,552){\rotatebox{-270}{\makebox(0,0){per command}}}
\put(1219,46){\makebox(0,0){Number of processes}}
\end{picture}
    }
    \caption{Comparison of the average number of reordering per command of \fbfs and \ffair  depending on the number of processes.}
    
    \label{fig:stabilityvsprocessesnosem}
\end{figure}

We measured the computation delays incurred by our reconciliation functions on different DAG sizes.
The results are presented in Figure~\ref{fig:computationvscommandsnosem} and we can see that stronger guarantees resulted in longer execution duration: fairness (\ffair) needed more computations than just stability (\fbfs).
On $1000$ commands, \ffair ran for $0.042$ seconds, roughly 20\% slower than \fbfs ($0.035$ seconds).
We also observed that \ffair induces less overhead compared to \fbfs as the number of commands grows: for $100$ commands, \ffair is $\approx 273\%$ times slower than \fbfs, but only $\approx 21\%$ slower on $1000$ commands.

\begin{figure}
    \centerline{
\setlength{\unitlength}{0.120450pt}
\ifx\plotpoint\undefined\newsavebox{\plotpoint}\fi
\ifx\transparent\undefined%
    \providecommand{\gpopaque}{}%
    \providecommand{\gptransparent}[2]{\color{.!#2}}%
\else%
    \providecommand{\gpopaque}{\transparent{1.0}}%
    \providecommand{\gptransparent}[2]{\transparent{#1}}%
\fi%
\begin{picture}(2159,1079)(0,0)
\miterjoin\buttcap
\color[rgb]{0.50,0.50,0.50}
\sbox{\plotpoint}{\rule[-0.200pt]{0.400pt}{0.400pt}}%
\linethickness{0.4pt}%
\multiput(313,277)(24.907,0.000){70}{\usebox{\plotpoint}}
\put(2033,277){\usebox{\plotpoint}}
\sbox{\plotpoint}{\rule[-0.600pt]{1.200pt}{1.200pt}}%
\linethickness{1.2pt}%
\Line(313,277)(354,277)
\Line(2033,277)(1992,277)
\put(280,277){\makebox(0,0)[r]{$0$}}
\sbox{\plotpoint}{\rule[-0.200pt]{0.400pt}{0.400pt}}%
\linethickness{0.4pt}%
\multiput(313,359)(24.907,0.000){70}{\usebox{\plotpoint}}
\put(2033,359){\usebox{\plotpoint}}
\sbox{\plotpoint}{\rule[-0.600pt]{1.200pt}{1.200pt}}%
\linethickness{1.2pt}%
\Line(313,359)(354,359)
\Line(2033,359)(1992,359)
\put(280,359){\makebox(0,0)[r]{$5000$}}
\sbox{\plotpoint}{\rule[-0.200pt]{0.400pt}{0.400pt}}%
\linethickness{0.4pt}%
\multiput(313,440)(24.907,0.000){70}{\usebox{\plotpoint}}
\put(2033,440){\usebox{\plotpoint}}
\sbox{\plotpoint}{\rule[-0.600pt]{1.200pt}{1.200pt}}%
\linethickness{1.2pt}%
\Line(313,440)(354,440)
\Line(2033,440)(1992,440)
\put(280,440){\makebox(0,0)[r]{$10000$}}
\sbox{\plotpoint}{\rule[-0.200pt]{0.400pt}{0.400pt}}%
\linethickness{0.4pt}%
\multiput(313,522)(24.907,0.000){70}{\usebox{\plotpoint}}
\put(2033,522){\usebox{\plotpoint}}
\sbox{\plotpoint}{\rule[-0.600pt]{1.200pt}{1.200pt}}%
\linethickness{1.2pt}%
\Line(313,522)(354,522)
\Line(2033,522)(1992,522)
\put(280,522){\makebox(0,0)[r]{$15000$}}
\sbox{\plotpoint}{\rule[-0.200pt]{0.400pt}{0.400pt}}%
\linethickness{0.4pt}%
\multiput(313,604)(24.907,0.000){70}{\usebox{\plotpoint}}
\put(2033,604){\usebox{\plotpoint}}
\sbox{\plotpoint}{\rule[-0.600pt]{1.200pt}{1.200pt}}%
\linethickness{1.2pt}%
\Line(313,604)(354,604)
\Line(2033,604)(1992,604)
\put(280,604){\makebox(0,0)[r]{$20000$}}
\sbox{\plotpoint}{\rule[-0.200pt]{0.400pt}{0.400pt}}%
\linethickness{0.4pt}%
\multiput(313,685)(24.907,0.000){70}{\usebox{\plotpoint}}
\put(2033,685){\usebox{\plotpoint}}
\sbox{\plotpoint}{\rule[-0.600pt]{1.200pt}{1.200pt}}%
\linethickness{1.2pt}%
\Line(313,685)(354,685)
\Line(2033,685)(1992,685)
\put(280,685){\makebox(0,0)[r]{$25000$}}
\sbox{\plotpoint}{\rule[-0.200pt]{0.400pt}{0.400pt}}%
\linethickness{0.4pt}%
\multiput(313,767)(24.907,0.000){70}{\usebox{\plotpoint}}
\put(2033,767){\usebox{\plotpoint}}
\sbox{\plotpoint}{\rule[-0.600pt]{1.200pt}{1.200pt}}%
\linethickness{1.2pt}%
\Line(313,767)(354,767)
\Line(2033,767)(1992,767)
\put(280,767){\makebox(0,0)[r]{$30000$}}
\sbox{\plotpoint}{\rule[-0.200pt]{0.400pt}{0.400pt}}%
\linethickness{0.4pt}%
\multiput(313,849)(24.907,0.000){2}{\usebox{\plotpoint}}
\put(346,849){\usebox{\plotpoint}}
\multiput(849,849)(24.907,0.000){48}{\usebox{\plotpoint}}
\put(2033,849){\usebox{\plotpoint}}
\sbox{\plotpoint}{\rule[-0.600pt]{1.200pt}{1.200pt}}%
\linethickness{1.2pt}%
\Line(313,849)(354,849)
\Line(2033,849)(1992,849)
\put(280,849){\makebox(0,0)[r]{$35000$}}
\sbox{\plotpoint}{\rule[-0.200pt]{0.400pt}{0.400pt}}%
\linethickness{0.4pt}%
\multiput(313,930)(24.907,0.000){2}{\usebox{\plotpoint}}
\put(346,930){\usebox{\plotpoint}}
\multiput(849,930)(24.907,0.000){48}{\usebox{\plotpoint}}
\put(2033,930){\usebox{\plotpoint}}
\sbox{\plotpoint}{\rule[-0.600pt]{1.200pt}{1.200pt}}%
\linethickness{1.2pt}%
\Line(313,930)(354,930)
\Line(2033,930)(1992,930)
\put(280,930){\makebox(0,0)[r]{$40000$}}
\sbox{\plotpoint}{\rule[-0.200pt]{0.400pt}{0.400pt}}%
\linethickness{0.4pt}%
\multiput(313,1012)(24.907,0.000){70}{\usebox{\plotpoint}}
\put(2033,1012){\usebox{\plotpoint}}
\sbox{\plotpoint}{\rule[-0.600pt]{1.200pt}{1.200pt}}%
\linethickness{1.2pt}%
\Line(313,1012)(354,1012)
\Line(2033,1012)(1992,1012)
\put(280,1012){\makebox(0,0)[r]{$45000$}}
\sbox{\plotpoint}{\rule[-0.200pt]{0.400pt}{0.400pt}}%
\linethickness{0.4pt}%
\multiput(313,277)(0.000,24.907){30}{\usebox{\plotpoint}}
\put(313,1012){\usebox{\plotpoint}}
\sbox{\plotpoint}{\rule[-0.600pt]{1.200pt}{1.200pt}}%
\linethickness{1.2pt}%
\Line(313,277)(313,318)
\Line(313,1012)(313,971)
\put(214,152){\rotatebox{45}{\makebox(0,0)[l]{10}}}
\sbox{\plotpoint}{\rule[-0.200pt]{0.400pt}{0.400pt}}%
\linethickness{0.4pt}%
\multiput(886,277)(0.000,24.907){30}{\usebox{\plotpoint}}
\put(886,1012){\usebox{\plotpoint}}
\sbox{\plotpoint}{\rule[-0.600pt]{1.200pt}{1.200pt}}%
\linethickness{1.2pt}%
\Line(886,277)(886,318)
\Line(886,1012)(886,971)
\put(787,152){\rotatebox{45}{\makebox(0,0)[l]{100}}}
\sbox{\plotpoint}{\rule[-0.200pt]{0.400pt}{0.400pt}}%
\linethickness{0.4pt}%
\multiput(1460,277)(0.000,24.907){30}{\usebox{\plotpoint}}
\put(1460,1012){\usebox{\plotpoint}}
\sbox{\plotpoint}{\rule[-0.600pt]{1.200pt}{1.200pt}}%
\linethickness{1.2pt}%
\Line(1460,277)(1460,318)
\Line(1460,1012)(1460,971)
\put(1361,152){\rotatebox{45}{\makebox(0,0)[l]{500}}}
\sbox{\plotpoint}{\rule[-0.200pt]{0.400pt}{0.400pt}}%
\linethickness{0.4pt}%
\multiput(2033,277)(0.000,24.907){30}{\usebox{\plotpoint}}
\put(2033,1012){\usebox{\plotpoint}}
\sbox{\plotpoint}{\rule[-0.600pt]{1.200pt}{1.200pt}}%
\linethickness{1.2pt}%
\Line(2033,277)(2033,318)
\Line(2033,1012)(2033,971)
\put(1934,152){\rotatebox{45}{\makebox(0,0)[l]{1000}}}
\polygon(313,1012)(313,277)(2033,277)(2033,1012)
\color{black}
\put(610,938){\makebox(0,0)[r]{\fbfs}}
\color[rgb]{0.22,0.49,0.72}
\sbox{\plotpoint}{\rule[-0.300pt]{0.600pt}{0.600pt}}%
\linethickness{0.6pt}%
\Line(643,938)(816,938)
\polyline(313,278)(313,278)(886,285)(1460,406)(2033,846)
\put(313,278){\makebox(0,0){$\ast$}}
\put(886,285){\makebox(0,0){$\ast$}}
\put(1460,406){\makebox(0,0){$\ast$}}
\put(2033,846){\makebox(0,0){$\ast$}}
\put(729,938){\makebox(0,0){$\ast$}}
\color{black}
\put(610,872){\makebox(0,0)[r]{\ffair}}
\color[rgb]{0.30,0.69,0.29}
\Line(643,872)(816,872)
\polyline(313,280)(313,280)(886,308)(1460,512)(2033,968)
\put(313,280){\makebox(0,0){$\ast$}}
\put(886,308){\makebox(0,0){$\ast$}}
\put(1460,512){\makebox(0,0){$\ast$}}
\put(2033,968){\makebox(0,0){$\ast$}}
\put(729,872){\makebox(0,0){$\ast$}}
\color{black}
\put(8,644){\rotatebox{-270}{\makebox(0,0){Average read duration ($\mu s$)}}}
\put(1173,46){\makebox(0,0){State size (number of commands)}}
\end{picture}
    }
    \caption{Comparison of the execution duration of \fbfs and \ffair, depending on size of the DAG (total number of commands).}
    
    \label{fig:computationvscommandsnosem}
\end{figure}

\subparagraph*{Fairness Under Partitions.}
In these experiments, we focus on the commands that stabilized with their initial contexts and referred to them as \emph{fairly stabilized commands}.
We also considered, for each process, the number of fairly stabilized commands they issued and measured fairness as the ratio between the \emph{minimum} and the \emph{maximum} of these counts across \emph{all processes}, called the \emph{fairness ratio}.

In a first experiment, we measured the proportion of fairly stabilized commands over the total number of commands issued by $16$ processes, all of which were temporarily isolated during a fraction of the execution.
For an execution lasting $d = 300$ seconds and a partition ratio $r$, the execution proceeds in three phases: processes start and issue commands as usual during the first $(1-r)d/2$ seconds; they are then isolated for $r d$ seconds while continuing to issue commands; and finally, connectivity is restored during the last $(1-r)d/2$ seconds.
Figure~\ref{fig:fairnessvspartitionsnosem} shows that we observed a proportion of fairly stabilized commands for \ffair around twice higher than for \fbfs.
This proportion depends on how quickly processes are able to communicate with each other, and decreases as partitions last longer.
Regardless of the partition duration, we observed that no process starved with \ffair, i.e., each process issued a command that fairly stabilized.
In contrast, we observed starvation for \fbfs even without partition.

\begin{figure}
    \centerline{
\setlength{\unitlength}{0.120450pt}
\ifx\plotpoint\undefined\newsavebox{\plotpoint}\fi
\ifx\transparent\undefined%
    \providecommand{\gpopaque}{}%
    \providecommand{\gptransparent}[2]{\color{.!#2}}%
\else%
    \providecommand{\gpopaque}{\transparent{1.0}}%
    \providecommand{\gptransparent}[2]{\transparent{#1}}%
\fi%
\begin{picture}(2159,960)(0,0)
\miterjoin\buttcap
\color[rgb]{0.50,0.50,0.50}
\sbox{\plotpoint}{\rule[-0.200pt]{0.400pt}{0.400pt}}%
\linethickness{0.4pt}%
\multiput(214,211)(24.907,0.000){75}{\usebox{\plotpoint}}
\put(2059,211){\usebox{\plotpoint}}
\sbox{\plotpoint}{\rule[-0.600pt]{1.200pt}{1.200pt}}%
\linethickness{1.2pt}%
\Line(214,211)(255,211)
\Line(2059,211)(2018,211)
\put(181,211){\makebox(0,0)[r]{$0$}}
\sbox{\plotpoint}{\rule[-0.200pt]{0.400pt}{0.400pt}}%
\linethickness{0.4pt}%
\multiput(214,314)(24.907,0.000){75}{\usebox{\plotpoint}}
\put(2059,314){\usebox{\plotpoint}}
\sbox{\plotpoint}{\rule[-0.600pt]{1.200pt}{1.200pt}}%
\linethickness{1.2pt}%
\Line(214,314)(255,314)
\Line(2059,314)(2018,314)
\put(181,314){\makebox(0,0)[r]{$2$}}
\sbox{\plotpoint}{\rule[-0.200pt]{0.400pt}{0.400pt}}%
\linethickness{0.4pt}%
\multiput(214,416)(24.907,0.000){75}{\usebox{\plotpoint}}
\put(2059,416){\usebox{\plotpoint}}
\sbox{\plotpoint}{\rule[-0.600pt]{1.200pt}{1.200pt}}%
\linethickness{1.2pt}%
\Line(214,416)(255,416)
\Line(2059,416)(2018,416)
\put(181,416){\makebox(0,0)[r]{$4$}}
\sbox{\plotpoint}{\rule[-0.200pt]{0.400pt}{0.400pt}}%
\linethickness{0.4pt}%
\multiput(214,519)(24.907,0.000){75}{\usebox{\plotpoint}}
\put(2059,519){\usebox{\plotpoint}}
\sbox{\plotpoint}{\rule[-0.600pt]{1.200pt}{1.200pt}}%
\linethickness{1.2pt}%
\Line(214,519)(255,519)
\Line(2059,519)(2018,519)
\put(181,519){\makebox(0,0)[r]{$6$}}
\sbox{\plotpoint}{\rule[-0.200pt]{0.400pt}{0.400pt}}%
\linethickness{0.4pt}%
\multiput(214,622)(24.907,0.000){75}{\usebox{\plotpoint}}
\put(2059,622){\usebox{\plotpoint}}
\sbox{\plotpoint}{\rule[-0.600pt]{1.200pt}{1.200pt}}%
\linethickness{1.2pt}%
\Line(214,622)(255,622)
\Line(2059,622)(2018,622)
\put(181,622){\makebox(0,0)[r]{$8$}}
\sbox{\plotpoint}{\rule[-0.200pt]{0.400pt}{0.400pt}}%
\linethickness{0.4pt}%
\multiput(214,724)(24.907,0.000){75}{\usebox{\plotpoint}}
\put(2059,724){\usebox{\plotpoint}}
\sbox{\plotpoint}{\rule[-0.600pt]{1.200pt}{1.200pt}}%
\linethickness{1.2pt}%
\Line(214,724)(255,724)
\Line(2059,724)(2018,724)
\put(181,724){\makebox(0,0)[r]{$10$}}
\sbox{\plotpoint}{\rule[-0.200pt]{0.400pt}{0.400pt}}%
\linethickness{0.4pt}%
\multiput(214,827)(24.907,0.000){75}{\usebox{\plotpoint}}
\put(2059,827){\usebox{\plotpoint}}
\sbox{\plotpoint}{\rule[-0.600pt]{1.200pt}{1.200pt}}%
\linethickness{1.2pt}%
\Line(214,827)(255,827)
\Line(2059,827)(2018,827)
\put(181,827){\makebox(0,0)[r]{$12$}}
\sbox{\plotpoint}{\rule[-0.200pt]{0.400pt}{0.400pt}}%
\linethickness{0.4pt}%
\multiput(583,211)(0.000,24.907){25}{\usebox{\plotpoint}}
\put(583,827){\usebox{\plotpoint}}
\sbox{\plotpoint}{\rule[-0.600pt]{1.200pt}{1.200pt}}%
\linethickness{1.2pt}%
\Line(583,211)(583,252)
\Line(583,827)(583,786)
\put(583,145){\makebox(0,0){0}}
\sbox{\plotpoint}{\rule[-0.200pt]{0.400pt}{0.400pt}}%
\linethickness{0.4pt}%
\multiput(952,211)(0.000,24.907){25}{\usebox{\plotpoint}}
\put(952,827){\usebox{\plotpoint}}
\sbox{\plotpoint}{\rule[-0.600pt]{1.200pt}{1.200pt}}%
\linethickness{1.2pt}%
\Line(952,211)(952,252)
\Line(952,827)(952,786)
\put(952,145){\makebox(0,0){0.1}}
\sbox{\plotpoint}{\rule[-0.200pt]{0.400pt}{0.400pt}}%
\linethickness{0.4pt}%
\multiput(1321,211)(0.000,24.907){25}{\usebox{\plotpoint}}
\put(1321,827){\usebox{\plotpoint}}
\sbox{\plotpoint}{\rule[-0.600pt]{1.200pt}{1.200pt}}%
\linethickness{1.2pt}%
\Line(1321,211)(1321,252)
\Line(1321,827)(1321,786)
\put(1321,145){\makebox(0,0){0.25}}
\sbox{\plotpoint}{\rule[-0.200pt]{0.400pt}{0.400pt}}%
\linethickness{0.4pt}%
\multiput(1690,211)(0.000,24.907){25}{\usebox{\plotpoint}}
\put(1690,827){\usebox{\plotpoint}}
\sbox{\plotpoint}{\rule[-0.600pt]{1.200pt}{1.200pt}}%
\linethickness{1.2pt}%
\Line(1690,211)(1690,252)
\Line(1690,827)(1690,786)
\put(1690,145){\makebox(0,0){0.5}}
\polygon(214,827)(214,211)(2059,211)(2059,827)
\color{black}
\put(897,886){\makebox(0,0)[r]{\fbfs}}
\color[rgb]{0.22,0.49,0.72}
\polygon*(930,870)(1103,870)(1103,903)(930,903)
\color{black}
\sbox{\plotpoint}{\rule[-0.200pt]{0.400pt}{0.400pt}}%
\linethickness{0.4pt}%
\Line(930,870)(1103,870)
\Line(1103,870)(1103,902)
\Line(1103,902)(930,902)
\Line(930,902)(930,870)
\color[rgb]{0.22,0.49,0.72}
\polygon*(495,211)(579,211)(579,502)(495,502)
\color{black}
\polygon(495,211)(495,501)(578,501)(578,211)
\color[rgb]{0.22,0.49,0.72}
\polygon*(864,211)(948,211)(948,444)(864,444)
\color{black}
\polygon(864,211)(864,443)(947,443)(947,211)
\color[rgb]{0.22,0.49,0.72}
\polygon*(1233,211)(1317,211)(1317,401)(1233,401)
\color{black}
\polygon(1233,211)(1233,400)(1316,400)(1316,211)
\color[rgb]{0.22,0.49,0.72}
\polygon*(1602,211)(1686,211)(1686,318)(1602,318)
\color{black}
\polygon(1602,211)(1602,317)(1685,317)(1685,211)
\put(1400,886){\makebox(0,0)[r]{\ffair}}
\color[rgb]{0.30,0.69,0.29}
\polygon*(1433,870)(1606,870)(1606,903)(1433,903)
\color{black}
\Line(1433,870)(1606,870)
\Line(1606,870)(1606,902)
\Line(1606,902)(1433,902)
\Line(1433,902)(1433,870)
\color[rgb]{0.30,0.69,0.29}
\polygon*(588,211)(672,211)(672,780)(588,780)
\color{black}
\polygon(588,211)(588,779)(671,779)(671,211)
\color[rgb]{0.30,0.69,0.29}
\polygon*(957,211)(1041,211)(1041,595)(957,595)
\color{black}
\polygon(957,211)(957,594)(1040,594)(1040,211)
\color[rgb]{0.30,0.69,0.29}
\polygon*(1326,211)(1410,211)(1410,579)(1326,579)
\color{black}
\polygon(1326,211)(1326,578)(1409,578)(1409,211)
\color[rgb]{0.30,0.69,0.29}
\polygon*(1695,211)(1779,211)(1779,447)(1695,447)
\color{black}
\polygon(1695,211)(1695,446)(1778,446)(1778,211)
\put(57,519){\rotatebox{-270}{\makebox(0,0){Fairly stabilized commands (\%)}}}
\put(1136,46){\makebox(0,0){Partition duration (ratio)}}
\end{picture}
    }
    \caption{Comparison of the number of fairly stabilized commands (\%) of \fbfs and \ffair, depending on how long processes were isolated.}
    
    \label{fig:fairnessvspartitionsnosem}
\end{figure}

\begin{figure}
    \centerline{
\setlength{\unitlength}{0.120450pt}
\ifx\plotpoint\undefined\newsavebox{\plotpoint}\fi
\ifx\transparent\undefined%
    \providecommand{\gpopaque}{}%
    \providecommand{\gptransparent}[2]{\color{.!#2}}%
\else%
    \providecommand{\gpopaque}{\transparent{1.0}}%
    \providecommand{\gptransparent}[2]{\transparent{#1}}%
\fi%
\begin{picture}(2159,839)(0,0)
\miterjoin\buttcap
\color[rgb]{0.50,0.50,0.50}
\sbox{\plotpoint}{\rule[-0.200pt]{0.400pt}{0.400pt}}%
\linethickness{0.4pt}%
\multiput(313,211)(24.907,0.000){71}{\usebox{\plotpoint}}
\put(2059,211){\usebox{\plotpoint}}
\sbox{\plotpoint}{\rule[-0.600pt]{1.200pt}{1.200pt}}%
\linethickness{1.2pt}%
\Line(313,211)(354,211)
\Line(2059,211)(2018,211)
\put(280,211){\makebox(0,0)[r]{$0$}}
\sbox{\plotpoint}{\rule[-0.200pt]{0.400pt}{0.400pt}}%
\linethickness{0.4pt}%
\multiput(313,323)(24.907,0.000){71}{\usebox{\plotpoint}}
\put(2059,323){\usebox{\plotpoint}}
\sbox{\plotpoint}{\rule[-0.600pt]{1.200pt}{1.200pt}}%
\linethickness{1.2pt}%
\Line(313,323)(354,323)
\Line(2059,323)(2018,323)
\put(280,323){\makebox(0,0)[r]{$0.2$}}
\sbox{\plotpoint}{\rule[-0.200pt]{0.400pt}{0.400pt}}%
\linethickness{0.4pt}%
\multiput(313,435)(24.907,0.000){71}{\usebox{\plotpoint}}
\put(2059,435){\usebox{\plotpoint}}
\sbox{\plotpoint}{\rule[-0.600pt]{1.200pt}{1.200pt}}%
\linethickness{1.2pt}%
\Line(313,435)(354,435)
\Line(2059,435)(2018,435)
\put(280,435){\makebox(0,0)[r]{$0.4$}}
\sbox{\plotpoint}{\rule[-0.200pt]{0.400pt}{0.400pt}}%
\linethickness{0.4pt}%
\multiput(313,548)(24.907,0.000){71}{\usebox{\plotpoint}}
\put(2059,548){\usebox{\plotpoint}}
\sbox{\plotpoint}{\rule[-0.600pt]{1.200pt}{1.200pt}}%
\linethickness{1.2pt}%
\Line(313,548)(354,548)
\Line(2059,548)(2018,548)
\put(280,548){\makebox(0,0)[r]{$0.6$}}
\sbox{\plotpoint}{\rule[-0.200pt]{0.400pt}{0.400pt}}%
\linethickness{0.4pt}%
\multiput(313,660)(24.907,0.000){71}{\usebox{\plotpoint}}
\put(2059,660){\usebox{\plotpoint}}
\sbox{\plotpoint}{\rule[-0.600pt]{1.200pt}{1.200pt}}%
\linethickness{1.2pt}%
\Line(313,660)(354,660)
\Line(2059,660)(2018,660)
\put(280,660){\makebox(0,0)[r]{$0.8$}}
\sbox{\plotpoint}{\rule[-0.200pt]{0.400pt}{0.400pt}}%
\linethickness{0.4pt}%
\multiput(313,772)(24.907,0.000){71}{\usebox{\plotpoint}}
\put(2059,772){\usebox{\plotpoint}}
\sbox{\plotpoint}{\rule[-0.600pt]{1.200pt}{1.200pt}}%
\linethickness{1.2pt}%
\Line(313,772)(354,772)
\Line(2059,772)(2018,772)
\put(280,772){\makebox(0,0)[r]{$1$}}
\sbox{\plotpoint}{\rule[-0.200pt]{0.400pt}{0.400pt}}%
\linethickness{0.4pt}%
\multiput(662,211)(0.000,24.907){23}{\usebox{\plotpoint}}
\put(662,772){\usebox{\plotpoint}}
\sbox{\plotpoint}{\rule[-0.600pt]{1.200pt}{1.200pt}}%
\linethickness{1.2pt}%
\Line(662,211)(662,252)
\Line(662,772)(662,731)
\put(662,145){\makebox(0,0){4}}
\sbox{\plotpoint}{\rule[-0.200pt]{0.400pt}{0.400pt}}%
\linethickness{0.4pt}%
\multiput(1011,211)(0.000,24.907){19}{\usebox{\plotpoint}}
\put(1011,665){\usebox{\plotpoint}}
\multiput(1011,731)(0.000,24.907){2}{\usebox{\plotpoint}}
\put(1011,772){\usebox{\plotpoint}}
\sbox{\plotpoint}{\rule[-0.600pt]{1.200pt}{1.200pt}}%
\linethickness{1.2pt}%
\Line(1011,211)(1011,252)
\Line(1011,772)(1011,731)
\put(1011,145){\makebox(0,0){8}}
\sbox{\plotpoint}{\rule[-0.200pt]{0.400pt}{0.400pt}}%
\linethickness{0.4pt}%
\multiput(1361,211)(0.000,24.907){19}{\usebox{\plotpoint}}
\put(1361,665){\usebox{\plotpoint}}
\multiput(1361,731)(0.000,24.907){2}{\usebox{\plotpoint}}
\put(1361,772){\usebox{\plotpoint}}
\sbox{\plotpoint}{\rule[-0.600pt]{1.200pt}{1.200pt}}%
\linethickness{1.2pt}%
\Line(1361,211)(1361,252)
\Line(1361,772)(1361,731)
\put(1361,145){\makebox(0,0){12}}
\sbox{\plotpoint}{\rule[-0.200pt]{0.400pt}{0.400pt}}%
\linethickness{0.4pt}%
\multiput(1710,211)(0.000,24.907){23}{\usebox{\plotpoint}}
\put(1710,772){\usebox{\plotpoint}}
\sbox{\plotpoint}{\rule[-0.600pt]{1.200pt}{1.200pt}}%
\linethickness{1.2pt}%
\Line(1710,211)(1710,252)
\Line(1710,772)(1710,731)
\put(1710,145){\makebox(0,0){16}}
\polygon(313,772)(313,211)(2059,211)(2059,772)
\color{black}
\put(947,698){\makebox(0,0)[r]{\fbfs}}
\color[rgb]{0.22,0.49,0.72}
\polygon*(980,682)(1153,682)(1153,715)(980,715)
\color{black}
\sbox{\plotpoint}{\rule[-0.200pt]{0.400pt}{0.400pt}}%
\linethickness{0.4pt}%
\Line(980,682)(1153,682)
\Line(1153,682)(1153,714)
\Line(1153,714)(980,714)
\Line(980,714)(980,682)
\color[rgb]{0.22,0.49,0.72}
\polygon*(579,211)(659,211)(659,372)(579,372)
\color{black}
\polygon(579,211)(579,371)(658,371)(658,211)
\color[rgb]{0.22,0.49,0.72}
\polygon*(928,211)(1008,211)(1008,279)(928,279)
\color{black}
\polygon(928,211)(928,278)(1007,278)(1007,211)
\put(1450,698){\makebox(0,0)[r]{\ffair}}
\color[rgb]{0.30,0.69,0.29}
\polygon*(1483,682)(1656,682)(1656,715)(1483,715)
\color{black}
\Line(1483,682)(1656,682)
\Line(1656,682)(1656,714)
\Line(1656,714)(1483,714)
\Line(1483,714)(1483,682)
\color[rgb]{0.30,0.69,0.29}
\polygon*(667,211)(746,211)(746,514)(667,514)
\color{black}
\polygon(667,211)(667,513)(745,513)(745,211)
\color[rgb]{0.30,0.69,0.29}
\polygon*(1016,211)(1095,211)(1095,536)(1016,536)
\color{black}
\polygon(1016,211)(1016,535)(1094,535)(1094,211)
\color[rgb]{0.30,0.69,0.29}
\polygon*(1365,211)(1445,211)(1445,532)(1365,532)
\color{black}
\polygon(1365,211)(1365,531)(1444,531)(1444,211)
\color[rgb]{0.30,0.69,0.29}
\polygon*(1714,211)(1794,211)(1794,493)(1714,493)
\color{black}
\polygon(1714,211)(1714,492)(1793,492)(1793,211)
\put(57,491){\rotatebox{-270}{\makebox(0,0){Fairness ratio of fairly}}}
\put(123,491){\rotatebox{-270}{\makebox(0,0){stabilized commands}}}
\put(1186,46){\makebox(0,0){Number of processes}}
\end{picture}
    }
    \caption{Comparison of the fairness ratio of \fbfs and \ffair, depending on the number of processes.}
    
    \label{fig:fairnessvsprocessesnosem}
\end{figure}
In a second experiment, we measured the fairness ratio of fairly stabilized commands during executions with a partition spanning one third of their duration, and we present the results when varying number of processes in Figure~\ref{fig:fairnessvsprocessesnosem}.
We observed that \fbfs exhibits less fairness than \ffair: for executions with respectively $4$ and $8$ processes, some processes registered a number of fairly stabilized commands respectively $3.5$ and $8$ times larger than others.
For larger systems, some processes starve with \fbfs. 
In contrast, \ffair never exhibits starvation and maintains a roughly constant fairness ratio across system sizes: for $4$ and $8$ processes, no process has more than $2.4$ times as many fairly stabilized commands as another, and for a system of size $16$, no process is favored by more than a factor of $2$.

\subsection{Set Data Type}\label{sec-a:set}
We conducted similar experiments but instantiating a set data type with \texttt{add} and \texttt{remove} operations, and comparing our reconciliation functions. 
Due to the semantics of the set, commands may now fail, e.g., adding an integer already present in the set or removing an integer not in the set.
A reordering does not necessarily imply a change in the outcome of a command, since commutative operations in its prefix may leave its result unchanged.
We adapt the evaluation metrics accordingly:
when measuring reorderings, we only consider those that effectively change the result of a command; for fairness, we consider goodput, i.e., commands that remain successful and whose result does not change throughout stabilization---referred as \emph{successful commands}.
  
For all the experiments, we considered the possible elements of the set to be the integers from $0$ to $9$, and the set initial value to be $\{ 0, 2, 4, 6, 8\}$.
Processes drawn uniformly whether to add or remove to the set and selected at random the integer to add or remove such that they always issued commands that were legal based on their local context.

\begin{figure}
    \centerline{
\setlength{\unitlength}{0.120450pt}
\ifx\plotpoint\undefined\newsavebox{\plotpoint}\fi
\ifx\transparent\undefined%
    \providecommand{\gpopaque}{}%
    \providecommand{\gptransparent}[2]{\color{.!#2}}%
\else%
    \providecommand{\gpopaque}{\transparent{1.0}}%
    \providecommand{\gptransparent}[2]{\transparent{#1}}%
\fi%
\begin{picture}(2159,960)(0,0)
\miterjoin\buttcap
\color[rgb]{0.50,0.50,0.50}
\sbox{\plotpoint}{\rule[-0.200pt]{0.400pt}{0.400pt}}%
\linethickness{0.4pt}%
\multiput(379,277)(24.907,0.000){68}{\usebox{\plotpoint}}
\put(2059,277){\usebox{\plotpoint}}
\sbox{\plotpoint}{\rule[-0.600pt]{1.200pt}{1.200pt}}%
\linethickness{1.2pt}%
\Line(379,277)(420,277)
\Line(2059,277)(2018,277)
\put(346,277){\makebox(0,0)[r]{$0.125$}}
\sbox{\plotpoint}{\rule[-0.200pt]{0.400pt}{0.400pt}}%
\linethickness{0.4pt}%
\multiput(379,482)(24.907,0.000){68}{\usebox{\plotpoint}}
\put(2059,482){\usebox{\plotpoint}}
\sbox{\plotpoint}{\rule[-0.600pt]{1.200pt}{1.200pt}}%
\linethickness{1.2pt}%
\Line(379,482)(420,482)
\Line(2059,482)(2018,482)
\put(346,482){\makebox(0,0)[r]{$0.25$}}
\sbox{\plotpoint}{\rule[-0.200pt]{0.400pt}{0.400pt}}%
\linethickness{0.4pt}%
\multiput(379,688)(24.907,0.000){68}{\usebox{\plotpoint}}
\put(2059,688){\usebox{\plotpoint}}
\sbox{\plotpoint}{\rule[-0.600pt]{1.200pt}{1.200pt}}%
\linethickness{1.2pt}%
\Line(379,688)(420,688)
\Line(2059,688)(2018,688)
\put(346,688){\makebox(0,0)[r]{$0.5$}}
\sbox{\plotpoint}{\rule[-0.200pt]{0.400pt}{0.400pt}}%
\linethickness{0.4pt}%
\multiput(379,893)(24.907,0.000){68}{\usebox{\plotpoint}}
\put(2059,893){\usebox{\plotpoint}}
\sbox{\plotpoint}{\rule[-0.600pt]{1.200pt}{1.200pt}}%
\linethickness{1.2pt}%
\Line(379,893)(420,893)
\Line(2059,893)(2018,893)
\put(346,893){\makebox(0,0)[r]{$1$}}
\sbox{\plotpoint}{\rule[-0.200pt]{0.400pt}{0.400pt}}%
\linethickness{0.4pt}%
\multiput(715,211)(0.000,24.907){24}{\usebox{\plotpoint}}
\put(715,786){\usebox{\plotpoint}}
\multiput(715,852)(0.000,24.907){2}{\usebox{\plotpoint}}
\put(715,893){\usebox{\plotpoint}}
\sbox{\plotpoint}{\rule[-0.600pt]{1.200pt}{1.200pt}}%
\linethickness{1.2pt}%
\Line(715,211)(715,252)
\Line(715,893)(715,852)
\put(715,145){\makebox(0,0){4}}
\sbox{\plotpoint}{\rule[-0.200pt]{0.400pt}{0.400pt}}%
\linethickness{0.4pt}%
\multiput(1051,211)(0.000,24.907){24}{\usebox{\plotpoint}}
\put(1051,786){\usebox{\plotpoint}}
\multiput(1051,852)(0.000,24.907){2}{\usebox{\plotpoint}}
\put(1051,893){\usebox{\plotpoint}}
\sbox{\plotpoint}{\rule[-0.600pt]{1.200pt}{1.200pt}}%
\linethickness{1.2pt}%
\Line(1051,211)(1051,252)
\Line(1051,893)(1051,852)
\put(1051,145){\makebox(0,0){8}}
\sbox{\plotpoint}{\rule[-0.200pt]{0.400pt}{0.400pt}}%
\linethickness{0.4pt}%
\multiput(1387,211)(0.000,24.907){24}{\usebox{\plotpoint}}
\put(1387,786){\usebox{\plotpoint}}
\multiput(1387,852)(0.000,24.907){2}{\usebox{\plotpoint}}
\put(1387,893){\usebox{\plotpoint}}
\sbox{\plotpoint}{\rule[-0.600pt]{1.200pt}{1.200pt}}%
\linethickness{1.2pt}%
\Line(1387,211)(1387,252)
\Line(1387,893)(1387,852)
\put(1387,145){\makebox(0,0){12}}
\sbox{\plotpoint}{\rule[-0.200pt]{0.400pt}{0.400pt}}%
\linethickness{0.4pt}%
\multiput(1723,211)(0.000,24.907){28}{\usebox{\plotpoint}}
\put(1723,893){\usebox{\plotpoint}}
\sbox{\plotpoint}{\rule[-0.600pt]{1.200pt}{1.200pt}}%
\linethickness{1.2pt}%
\Line(1723,211)(1723,252)
\Line(1723,893)(1723,852)
\put(1723,145){\makebox(0,0){16}}
\polygon(379,893)(379,211)(2059,211)(2059,893)
\color{black}
\put(676,819){\makebox(0,0)[r]{\fbfs}}
\color[rgb]{0.22,0.49,0.72}
\polygon*(709,803)(882,803)(882,836)(709,836)
\color{black}
\sbox{\plotpoint}{\rule[-0.200pt]{0.400pt}{0.400pt}}%
\linethickness{0.4pt}%
\Line(709,803)(882,803)
\Line(882,803)(882,835)
\Line(882,835)(709,835)
\Line(709,835)(709,803)
\color[rgb]{0.22,0.49,0.72}
\polygon*(635,211)(712,211)(712,264)(635,264)
\color{black}
\polygon(635,211)(635,263)(711,263)(711,211)
\color[rgb]{0.22,0.49,0.72}
\polygon*(971,211)(1048,211)(1048,292)(971,292)
\color{black}
\polygon(971,211)(971,291)(1047,291)(1047,211)
\color[rgb]{0.22,0.49,0.72}
\polygon*(1307,211)(1384,211)(1384,351)(1307,351)
\color{black}
\polygon(1307,211)(1307,350)(1383,350)(1383,211)
\color[rgb]{0.22,0.49,0.72}
\polygon*(1643,211)(1720,211)(1720,396)(1643,396)
\color{black}
\polygon(1643,211)(1643,395)(1719,395)(1719,211)
\put(1179,819){\makebox(0,0)[r]{\ffair}}
\color[rgb]{0.30,0.69,0.29}
\polygon*(1212,803)(1385,803)(1385,836)(1212,836)
\color{black}
\Line(1212,803)(1385,803)
\Line(1385,803)(1385,835)
\Line(1385,835)(1212,835)
\Line(1212,835)(1212,803)
\color[rgb]{0.30,0.69,0.29}
\polygon*(719,211)(796,211)(796,483)(719,483)
\color{black}
\polygon(719,211)(719,482)(795,482)(795,211)
\color[rgb]{0.30,0.69,0.29}
\polygon*(1055,211)(1132,211)(1132,678)(1055,678)
\color{black}
\polygon(1055,211)(1055,677)(1131,677)(1131,211)
\color[rgb]{0.30,0.69,0.29}
\polygon*(1391,211)(1468,211)(1468,711)(1391,711)
\color{black}
\polygon(1391,211)(1391,710)(1467,710)(1467,211)
\color[rgb]{0.30,0.69,0.29}
\polygon*(1727,211)(1804,211)(1804,894)(1727,894)
\color{black}
\polygon(1727,211)(1727,893)(1803,893)(1803,211)
\put(24,552){\rotatebox{-270}{\makebox(0,0){Average number of change}}}
\put(90,552){\rotatebox{-270}{\makebox(0,0){of outcome per command}}}
\put(1219,46){\makebox(0,0){Number of processes}}
\end{picture}
    }
    \caption{Comparison of the average number of change of result per command of \fbfs and \ffair, depending on the number of processes.}
    
    \label{fig:stabilityvsprocessesremovewins}
\end{figure}

Regarding reorderings, Figure~\ref{fig:stabilityvsprocessesremovewins} shows that \ffair, as previously, changed the result more often than \fbfs in order to preserve fairness, with more reoderings as the number of processes increases.
Due to commutativity in the semantics, we observed less change of outcome than reorderings but the difference between \fbfs and \ffair in change of outcome was similar to their difference in reorderings.

\begin{figure}
    \centerline{
\setlength{\unitlength}{0.120450pt}
\ifx\plotpoint\undefined\newsavebox{\plotpoint}\fi
\ifx\transparent\undefined%
    \providecommand{\gpopaque}{}%
    \providecommand{\gptransparent}[2]{\color{.!#2}}%
\else%
    \providecommand{\gpopaque}{\transparent{1.0}}%
    \providecommand{\gptransparent}[2]{\transparent{#1}}%
\fi%
\begin{picture}(2159,960)(0,0)
\miterjoin\buttcap
\color[rgb]{0.50,0.50,0.50}
\sbox{\plotpoint}{\rule[-0.200pt]{0.400pt}{0.400pt}}%
\linethickness{0.4pt}%
\multiput(214,211)(24.907,0.000){75}{\usebox{\plotpoint}}
\put(2059,211){\usebox{\plotpoint}}
\sbox{\plotpoint}{\rule[-0.600pt]{1.200pt}{1.200pt}}%
\linethickness{1.2pt}%
\Line(214,211)(255,211)
\Line(2059,211)(2018,211)
\put(181,211){\makebox(0,0)[r]{$0$}}
\sbox{\plotpoint}{\rule[-0.200pt]{0.400pt}{0.400pt}}%
\linethickness{0.4pt}%
\multiput(214,288)(24.907,0.000){75}{\usebox{\plotpoint}}
\put(2059,288){\usebox{\plotpoint}}
\sbox{\plotpoint}{\rule[-0.600pt]{1.200pt}{1.200pt}}%
\linethickness{1.2pt}%
\Line(214,288)(255,288)
\Line(2059,288)(2018,288)
\put(181,288){\makebox(0,0)[r]{$10$}}
\sbox{\plotpoint}{\rule[-0.200pt]{0.400pt}{0.400pt}}%
\linethickness{0.4pt}%
\multiput(214,365)(24.907,0.000){75}{\usebox{\plotpoint}}
\put(2059,365){\usebox{\plotpoint}}
\sbox{\plotpoint}{\rule[-0.600pt]{1.200pt}{1.200pt}}%
\linethickness{1.2pt}%
\Line(214,365)(255,365)
\Line(2059,365)(2018,365)
\put(181,365){\makebox(0,0)[r]{$20$}}
\sbox{\plotpoint}{\rule[-0.200pt]{0.400pt}{0.400pt}}%
\linethickness{0.4pt}%
\multiput(214,442)(24.907,0.000){75}{\usebox{\plotpoint}}
\put(2059,442){\usebox{\plotpoint}}
\sbox{\plotpoint}{\rule[-0.600pt]{1.200pt}{1.200pt}}%
\linethickness{1.2pt}%
\Line(214,442)(255,442)
\Line(2059,442)(2018,442)
\put(181,442){\makebox(0,0)[r]{$30$}}
\sbox{\plotpoint}{\rule[-0.200pt]{0.400pt}{0.400pt}}%
\linethickness{0.4pt}%
\multiput(214,519)(24.907,0.000){75}{\usebox{\plotpoint}}
\put(2059,519){\usebox{\plotpoint}}
\sbox{\plotpoint}{\rule[-0.600pt]{1.200pt}{1.200pt}}%
\linethickness{1.2pt}%
\Line(214,519)(255,519)
\Line(2059,519)(2018,519)
\put(181,519){\makebox(0,0)[r]{$40$}}
\sbox{\plotpoint}{\rule[-0.200pt]{0.400pt}{0.400pt}}%
\linethickness{0.4pt}%
\multiput(214,596)(24.907,0.000){75}{\usebox{\plotpoint}}
\put(2059,596){\usebox{\plotpoint}}
\sbox{\plotpoint}{\rule[-0.600pt]{1.200pt}{1.200pt}}%
\linethickness{1.2pt}%
\Line(214,596)(255,596)
\Line(2059,596)(2018,596)
\put(181,596){\makebox(0,0)[r]{$50$}}
\sbox{\plotpoint}{\rule[-0.200pt]{0.400pt}{0.400pt}}%
\linethickness{0.4pt}%
\multiput(214,673)(24.907,0.000){75}{\usebox{\plotpoint}}
\put(2059,673){\usebox{\plotpoint}}
\sbox{\plotpoint}{\rule[-0.600pt]{1.200pt}{1.200pt}}%
\linethickness{1.2pt}%
\Line(214,673)(255,673)
\Line(2059,673)(2018,673)
\put(181,673){\makebox(0,0)[r]{$60$}}
\sbox{\plotpoint}{\rule[-0.200pt]{0.400pt}{0.400pt}}%
\linethickness{0.4pt}%
\multiput(214,750)(24.907,0.000){75}{\usebox{\plotpoint}}
\put(2059,750){\usebox{\plotpoint}}
\sbox{\plotpoint}{\rule[-0.600pt]{1.200pt}{1.200pt}}%
\linethickness{1.2pt}%
\Line(214,750)(255,750)
\Line(2059,750)(2018,750)
\put(181,750){\makebox(0,0)[r]{$70$}}
\sbox{\plotpoint}{\rule[-0.200pt]{0.400pt}{0.400pt}}%
\linethickness{0.4pt}%
\multiput(214,827)(24.907,0.000){75}{\usebox{\plotpoint}}
\put(2059,827){\usebox{\plotpoint}}
\sbox{\plotpoint}{\rule[-0.600pt]{1.200pt}{1.200pt}}%
\linethickness{1.2pt}%
\Line(214,827)(255,827)
\Line(2059,827)(2018,827)
\put(181,827){\makebox(0,0)[r]{$80$}}
\sbox{\plotpoint}{\rule[-0.200pt]{0.400pt}{0.400pt}}%
\linethickness{0.4pt}%
\multiput(583,211)(0.000,24.907){25}{\usebox{\plotpoint}}
\put(583,827){\usebox{\plotpoint}}
\sbox{\plotpoint}{\rule[-0.600pt]{1.200pt}{1.200pt}}%
\linethickness{1.2pt}%
\Line(583,211)(583,252)
\Line(583,827)(583,786)
\put(583,145){\makebox(0,0){0}}
\sbox{\plotpoint}{\rule[-0.200pt]{0.400pt}{0.400pt}}%
\linethickness{0.4pt}%
\multiput(952,211)(0.000,24.907){25}{\usebox{\plotpoint}}
\put(952,827){\usebox{\plotpoint}}
\sbox{\plotpoint}{\rule[-0.600pt]{1.200pt}{1.200pt}}%
\linethickness{1.2pt}%
\Line(952,211)(952,252)
\Line(952,827)(952,786)
\put(952,145){\makebox(0,0){0.1}}
\sbox{\plotpoint}{\rule[-0.200pt]{0.400pt}{0.400pt}}%
\linethickness{0.4pt}%
\multiput(1321,211)(0.000,24.907){25}{\usebox{\plotpoint}}
\put(1321,827){\usebox{\plotpoint}}
\sbox{\plotpoint}{\rule[-0.600pt]{1.200pt}{1.200pt}}%
\linethickness{1.2pt}%
\Line(1321,211)(1321,252)
\Line(1321,827)(1321,786)
\put(1321,145){\makebox(0,0){0.25}}
\sbox{\plotpoint}{\rule[-0.200pt]{0.400pt}{0.400pt}}%
\linethickness{0.4pt}%
\multiput(1690,211)(0.000,24.907){25}{\usebox{\plotpoint}}
\put(1690,827){\usebox{\plotpoint}}
\sbox{\plotpoint}{\rule[-0.600pt]{1.200pt}{1.200pt}}%
\linethickness{1.2pt}%
\Line(1690,211)(1690,252)
\Line(1690,827)(1690,786)
\put(1690,145){\makebox(0,0){0.5}}
\polygon(214,827)(214,211)(2059,211)(2059,827)
\color{black}
\put(897,886){\makebox(0,0)[r]{\fbfs}}
\color[rgb]{0.22,0.49,0.72}
\polygon*(930,870)(1103,870)(1103,903)(930,903)
\color{black}
\sbox{\plotpoint}{\rule[-0.200pt]{0.400pt}{0.400pt}}%
\linethickness{0.4pt}%
\Line(930,870)(1103,870)
\Line(1103,870)(1103,902)
\Line(1103,902)(930,902)
\Line(930,902)(930,870)
\color[rgb]{0.22,0.49,0.72}
\polygon*(495,211)(579,211)(579,731)(495,731)
\color{black}
\polygon(495,211)(495,730)(578,730)(578,211)
\color[rgb]{0.22,0.49,0.72}
\polygon*(864,211)(948,211)(948,704)(864,704)
\color{black}
\polygon(864,211)(864,703)(947,703)(947,211)
\color[rgb]{0.22,0.49,0.72}
\polygon*(1233,211)(1317,211)(1317,671)(1233,671)
\color{black}
\polygon(1233,211)(1233,670)(1316,670)(1316,211)
\color[rgb]{0.22,0.49,0.72}
\polygon*(1602,211)(1686,211)(1686,644)(1602,644)
\color{black}
\polygon(1602,211)(1602,643)(1685,643)(1685,211)
\put(1400,886){\makebox(0,0)[r]{\ffair}}
\color[rgb]{0.30,0.69,0.29}
\polygon*(1433,870)(1606,870)(1606,903)(1433,903)
\color{black}
\Line(1433,870)(1606,870)
\Line(1606,870)(1606,902)
\Line(1606,902)(1433,902)
\Line(1433,902)(1433,870)
\color[rgb]{0.30,0.69,0.29}
\polygon*(588,211)(672,211)(672,700)(588,700)
\color{black}
\polygon(588,211)(588,699)(671,699)(671,211)
\color[rgb]{0.30,0.69,0.29}
\polygon*(957,211)(1041,211)(1041,704)(957,704)
\color{black}
\polygon(957,211)(957,703)(1040,703)(1040,211)
\color[rgb]{0.30,0.69,0.29}
\polygon*(1326,211)(1410,211)(1410,718)(1326,718)
\color{black}
\polygon(1326,211)(1326,717)(1409,717)(1409,211)
\color[rgb]{0.30,0.69,0.29}
\polygon*(1695,211)(1779,211)(1779,786)(1695,786)
\color{black}
\polygon(1695,211)(1695,785)(1778,785)(1778,211)
\put(57,519){\rotatebox{-270}{\makebox(0,0){Successful commands (\%)}}}
\put(1136,46){\makebox(0,0){Partition duration (ratio)}}
\end{picture}
    }
    \caption{Comparison of the number of successful commands for \fbfs and \ffair, depending on how long processes were isolated.}
    
    \label{fig:fairnessvspartitionsremovewins}
\end{figure}
As for fairness, Figure~\ref{fig:fairnessvspartitionsremovewins} shows that without partition or with a short one ($0.1$), \fbfs yields a similar fraction of successful commands to the one of \ffair.
However, as partition duration increases, the fraction of successful commands under \fbfs steadily decreases, from $67\%$ for no partition to $56\%$ when partitions span half of the execution.
\ffair exhibits the opposite trend, with successful commands increasing from $64\%$ to $75\%$ as partitions last longer.
This behavior is expected: during partitions, processes issue sequences of successful commands that \ffair preserves upon reconciliation, whereas \fbfs may break these sequences by inserting concurrent commands at the same distance from the root.
We observed more successful than fairly stabilized commands due to commutativity allowing some commands, that would have been disabled otherwise, to still be successful despite reorderings.

\begin{figure}
    \centerline{
\setlength{\unitlength}{0.120450pt}
\ifx\plotpoint\undefined\newsavebox{\plotpoint}\fi
\ifx\transparent\undefined%
    \providecommand{\gpopaque}{}%
    \providecommand{\gptransparent}[2]{\color{.!#2}}%
\else%
    \providecommand{\gpopaque}{\transparent{1.0}}%
    \providecommand{\gptransparent}[2]{\transparent{#1}}%
\fi%
\begin{picture}(2159,839)(0,0)
\miterjoin\buttcap
\color[rgb]{0.50,0.50,0.50}
\sbox{\plotpoint}{\rule[-0.200pt]{0.400pt}{0.400pt}}%
\linethickness{0.4pt}%
\multiput(313,211)(24.907,0.000){71}{\usebox{\plotpoint}}
\put(2059,211){\usebox{\plotpoint}}
\sbox{\plotpoint}{\rule[-0.600pt]{1.200pt}{1.200pt}}%
\linethickness{1.2pt}%
\Line(313,211)(354,211)
\Line(2059,211)(2018,211)
\put(280,211){\makebox(0,0)[r]{$0$}}
\sbox{\plotpoint}{\rule[-0.200pt]{0.400pt}{0.400pt}}%
\linethickness{0.4pt}%
\multiput(313,323)(24.907,0.000){71}{\usebox{\plotpoint}}
\put(2059,323){\usebox{\plotpoint}}
\sbox{\plotpoint}{\rule[-0.600pt]{1.200pt}{1.200pt}}%
\linethickness{1.2pt}%
\Line(313,323)(354,323)
\Line(2059,323)(2018,323)
\put(280,323){\makebox(0,0)[r]{$0.2$}}
\sbox{\plotpoint}{\rule[-0.200pt]{0.400pt}{0.400pt}}%
\linethickness{0.4pt}%
\multiput(313,435)(24.907,0.000){71}{\usebox{\plotpoint}}
\put(2059,435){\usebox{\plotpoint}}
\sbox{\plotpoint}{\rule[-0.600pt]{1.200pt}{1.200pt}}%
\linethickness{1.2pt}%
\Line(313,435)(354,435)
\Line(2059,435)(2018,435)
\put(280,435){\makebox(0,0)[r]{$0.4$}}
\sbox{\plotpoint}{\rule[-0.200pt]{0.400pt}{0.400pt}}%
\linethickness{0.4pt}%
\multiput(313,548)(24.907,0.000){71}{\usebox{\plotpoint}}
\put(2059,548){\usebox{\plotpoint}}
\sbox{\plotpoint}{\rule[-0.600pt]{1.200pt}{1.200pt}}%
\linethickness{1.2pt}%
\Line(313,548)(354,548)
\Line(2059,548)(2018,548)
\put(280,548){\makebox(0,0)[r]{$0.6$}}
\sbox{\plotpoint}{\rule[-0.200pt]{0.400pt}{0.400pt}}%
\linethickness{0.4pt}%
\multiput(313,660)(24.907,0.000){71}{\usebox{\plotpoint}}
\put(2059,660){\usebox{\plotpoint}}
\sbox{\plotpoint}{\rule[-0.600pt]{1.200pt}{1.200pt}}%
\linethickness{1.2pt}%
\Line(313,660)(354,660)
\Line(2059,660)(2018,660)
\put(280,660){\makebox(0,0)[r]{$0.8$}}
\sbox{\plotpoint}{\rule[-0.200pt]{0.400pt}{0.400pt}}%
\linethickness{0.4pt}%
\multiput(313,772)(24.907,0.000){71}{\usebox{\plotpoint}}
\put(2059,772){\usebox{\plotpoint}}
\sbox{\plotpoint}{\rule[-0.600pt]{1.200pt}{1.200pt}}%
\linethickness{1.2pt}%
\Line(313,772)(354,772)
\Line(2059,772)(2018,772)
\put(280,772){\makebox(0,0)[r]{$1$}}
\sbox{\plotpoint}{\rule[-0.200pt]{0.400pt}{0.400pt}}%
\linethickness{0.4pt}%
\multiput(662,211)(0.000,24.907){23}{\usebox{\plotpoint}}
\put(662,772){\usebox{\plotpoint}}
\sbox{\plotpoint}{\rule[-0.600pt]{1.200pt}{1.200pt}}%
\linethickness{1.2pt}%
\Line(662,211)(662,252)
\Line(662,772)(662,731)
\put(662,145){\makebox(0,0){4}}
\sbox{\plotpoint}{\rule[-0.200pt]{0.400pt}{0.400pt}}%
\linethickness{0.4pt}%
\multiput(1011,211)(0.000,24.907){19}{\usebox{\plotpoint}}
\put(1011,665){\usebox{\plotpoint}}
\multiput(1011,731)(0.000,24.907){2}{\usebox{\plotpoint}}
\put(1011,772){\usebox{\plotpoint}}
\sbox{\plotpoint}{\rule[-0.600pt]{1.200pt}{1.200pt}}%
\linethickness{1.2pt}%
\Line(1011,211)(1011,252)
\Line(1011,772)(1011,731)
\put(1011,145){\makebox(0,0){8}}
\sbox{\plotpoint}{\rule[-0.200pt]{0.400pt}{0.400pt}}%
\linethickness{0.4pt}%
\multiput(1361,211)(0.000,24.907){19}{\usebox{\plotpoint}}
\put(1361,665){\usebox{\plotpoint}}
\multiput(1361,731)(0.000,24.907){2}{\usebox{\plotpoint}}
\put(1361,772){\usebox{\plotpoint}}
\sbox{\plotpoint}{\rule[-0.600pt]{1.200pt}{1.200pt}}%
\linethickness{1.2pt}%
\Line(1361,211)(1361,252)
\Line(1361,772)(1361,731)
\put(1361,145){\makebox(0,0){12}}
\sbox{\plotpoint}{\rule[-0.200pt]{0.400pt}{0.400pt}}%
\linethickness{0.4pt}%
\multiput(1710,211)(0.000,24.907){23}{\usebox{\plotpoint}}
\put(1710,772){\usebox{\plotpoint}}
\sbox{\plotpoint}{\rule[-0.600pt]{1.200pt}{1.200pt}}%
\linethickness{1.2pt}%
\Line(1710,211)(1710,252)
\Line(1710,772)(1710,731)
\put(1710,145){\makebox(0,0){16}}
\polygon(313,772)(313,211)(2059,211)(2059,772)
\color{black}
\put(947,698){\makebox(0,0)[r]{\fbfs}}
\color[rgb]{0.22,0.49,0.72}
\polygon*(980,682)(1153,682)(1153,715)(980,715)
\color{black}
\sbox{\plotpoint}{\rule[-0.200pt]{0.400pt}{0.400pt}}%
\linethickness{0.4pt}%
\Line(980,682)(1153,682)
\Line(1153,682)(1153,714)
\Line(1153,714)(980,714)
\Line(980,714)(980,682)
\color[rgb]{0.22,0.49,0.72}
\polygon*(579,211)(659,211)(659,361)(579,361)
\color{black}
\polygon(579,211)(579,360)(658,360)(658,211)
\color[rgb]{0.22,0.49,0.72}
\polygon*(928,211)(1008,211)(1008,351)(928,351)
\color{black}
\polygon(928,211)(928,350)(1007,350)(1007,211)
\color[rgb]{0.22,0.49,0.72}
\polygon*(1278,211)(1357,211)(1357,339)(1278,339)
\color{black}
\polygon(1278,211)(1278,338)(1356,338)(1356,211)
\color[rgb]{0.22,0.49,0.72}
\polygon*(1627,211)(1706,211)(1706,335)(1627,335)
\color{black}
\polygon(1627,211)(1627,334)(1705,334)(1705,211)
\put(1450,698){\makebox(0,0)[r]{\ffair}}
\color[rgb]{0.30,0.69,0.29}
\polygon*(1483,682)(1656,682)(1656,715)(1483,715)
\color{black}
\Line(1483,682)(1656,682)
\Line(1656,682)(1656,714)
\Line(1656,714)(1483,714)
\Line(1483,714)(1483,682)
\color[rgb]{0.30,0.69,0.29}
\polygon*(667,211)(746,211)(746,379)(667,379)
\color{black}
\polygon(667,211)(667,378)(745,378)(745,211)
\color[rgb]{0.30,0.69,0.29}
\polygon*(1016,211)(1095,211)(1095,389)(1016,389)
\color{black}
\polygon(1016,211)(1016,388)(1094,388)(1094,211)
\color[rgb]{0.30,0.69,0.29}
\polygon*(1365,211)(1445,211)(1445,387)(1365,387)
\color{black}
\polygon(1365,211)(1365,386)(1444,386)(1444,211)
\color[rgb]{0.30,0.69,0.29}
\polygon*(1714,211)(1794,211)(1794,423)(1714,423)
\color{black}
\polygon(1714,211)(1714,422)(1793,422)(1793,211)
\put(57,491){\rotatebox{-270}{\makebox(0,0){Fairness ratio of}}}
\put(123,491){\rotatebox{-270}{\makebox(0,0){successful commands}}}
\put(1186,46){\makebox(0,0){Number of processes}}
\end{picture}
    }
    \caption{Comparison of the fairness ratio of \fbfs and \ffair, depending on the number of processes.}
    
    \label{fig:fairnessvsprocessesremovewins}
\end{figure}
Figure~\ref{fig:fairnessvsprocessesremovewins} reports fairness when a partition spans one third of the execution.
For small systems ($n=4$), \fbfs and \ffair exhibit similar behavior, with the most favored process issuing about $3\times$ more successful commands than the least favored one.
As the number of processes increases, \fbfs becomes less and less fair, reaching a factor of $4.5$ for $n=12$ and $n=16$.
In contrast, \ffair improves with scale: for $n=16$, no process issues more than $2.7\times$ the number of successful commands of another.
Contrarily to without a specific data type, \fbfs exhibited no starvation, again due to commutativity.

\end{document}